\documentclass[times, nohelvet]{bioRxiv}
\usepackage{csquotes}
\usepackage{dsfont}
\usepackage{algorithm}
\usepackage[noend]{algpseudocode}
\usepackage{amsmath}
\usepackage{array}
\usepackage{bm}
\usepackage{lipsum}
\usepackage[export]{adjustbox}
\usepackage{subcaption}
\usepackage{varwidth}
\usepackage{booktabs}
\usepackage{longtable}
\usepackage{nameref}
\usepackage{textcomp}
\usepackage{setspace}
\usepackage[section]{placeins}

\usepackage{ulem}
\usepackage{changebar}

\newtoggle{draft}
\togglefalse{draft} 

\usepackage{xpatch}
\usepackage[backend=biber, style=authoryear, dashed=false]{biblatex}

\DeclareFieldFormat[article]{volume}{#1\addcolon\space}
\DeclareFieldFormat{pages}{#1}

\renewbibmacro{in:}{}

\makeatletter
\def\act@on@bibmacro#1#2{%
  \expandafter#1\csname abx@macro@\detokenize{#2}\endcsname
}
\def\patchbibmacro{\act@on@bibmacro\patchcmd}
\def\pretobibmacro{\act@on@bibmacro\pretocmd}
\def\apptobibmacro{\act@on@bibmacro\apptocmd}
\def\showbibmacro{\act@on@bibmacro\show}
\makeatother

\usepackage[frozencache=true,cachedir=minted-cache]{minted}

\DeclareRobustCommand{\sbseries}{\fontseries{sb}\selectfont}
\DeclareTextFontCommand{\textsb}{\sbseries}

\begin{document}

\onecolumn
\leadauthor{Penn}

\title{Leaping through tree space: continuous phylogenetic inference for rooted and unrooted trees}
\shorttitle{Leaping through tree space}

\author[1\textbf{*}]{Matthew J Penn}
\author[2\textbf{*}]{Neil Scheidwasser}
\author[6]{Joseph Penn}
\author[1, 4, 5]{Christl A Donnelly}
\author[3\textbf{*}]{David A Duchêne}
\author[2, 5 \Letter \textbf{*}]{Samir Bhatt}
\affil[1]{Department of Statistics, University of Oxford, Oxford, United Kingdom}
\affil[2]{Section of Epidemiology, University of Copenhagen, Copenhagen, Denmark}
\affil[3]{Center for Evolutionary Hologenomics, University of Copenhagen, Copenhagen, Denmark}
\affil[4]{Pandemic Sciences Institute, University of Oxford, Oxford, United Kingdom}
\affil[5]{MRC Centre for Global Infectious Disease Analysis, Department of Infectious Disease Epidemiology, School of Public Health, Faculty of Medicine, Imperial College London, London, United Kingdom}
\affil[6]{Department of Physics, University of Oxford, Oxford, United Kingdom}
\affil[*]{Equal contribution}
\date{}
\maketitle

\begin{abstract}
Phylogenetics is now fundamental in life sciences, providing insights into the earliest branches of life and the origins and spread of epidemics. However, finding suitable phylogenies from the vast space of possible trees remains challenging. To address this problem, for the first time, we perform both tree exploration and inference in a continuous space where the computation of gradients is possible. This continuous relaxation allows for major leaps across tree space in both rooted and unrooted trees, and is less susceptible to convergence to local minima. Our approach outperforms the current best methods for inference on unrooted trees and, in simulation, accurately infers the tree and root in ultrametric cases. The approach is effective in cases of empirical data with negligible amounts of data, which we demonstrate on the phylogeny of jawed vertebrates. Indeed, only a few genes with an ultrametric signal were generally sufficient for resolving the major lineages of vertebrates. Optimisation is possible via automatic differentiation and our method presents an effective way forwards for exploring the most difficult, data-deficient phylogenetic questions.

\end{abstract}


\begin{corrauthor}
\textcolor{red}{s.bhatt\at imperial.ac.uk}

\end{corrauthor}

\section*{Significance statement}
Phylogenetics is vital in life sciences, revealing early life origins and epidemic dynamics. A central challenge in inferring the best tree from a set of data is that exploring the vastness of tree space is computationally hard. Here we recast the exploration and inference problem from a discrete to a continuous space. Our method performs well on unrooted trees and can accurately infer both the tree and the root in ultrametric cases, but at a substantially increased computational cost. Our approach represents a shift in the methodology to explore tree space, and opens the possibilities of new efficient forms of inference.

\section*{Introduction}

Phylogenetic inference, the task of reconstructing the evolutionary relationships across taxonomic units given observational data, has a wide range of theoretical and practical applications in biology, such as evolution~\parencite{cavalli1967, felsenstein2004, omeara2012}, conservation~\parencite{rolland_cadotte} and epidemiology~\parencite{Grenfell2004-yl, faria2021}, but also in comparative linguistics~\parencite{mace2005} and cultural anthropology~\parencite{collard2006, morrison2014}. In particular, the COVID-19 pandemic has catalysed the development of efficient phylogenetic tools and methods to better understand the virus' origin, spread, and evolution~\parencite{lemoine2021, otoole2021, attwood2022, sanderson2022, turakhia2022, voznica2022, demaio2023}. For biological problems, tree inference is primarily informed by molecular sequence data (i.e., nucleotide or amino acid sequences), for which an extensive body of literature exists~\parencite{sanderson2002, yang2006, yang2012}. Other types of biological data such as morphology~\parencite{Lee2015-es}, fossils~\parencite{Morlon2011-pe}, and auditory communication in animals~\parencite{arato2021} can also be used as input.

Two key parameters considered when inferring a phylogenetic tree include the \emph{topology}, the branching pattern that specifies the evolutionary relationships between operational taxonomic units, and \emph{branch lengths}, the amount of evolutionary divergence that occurred between the branching events. A substantial amount of research has been conducted on how to parameterise branch lengths~\parencite{bromham2003, dosreis2016}, especially through the use of various molecular clocks~\parencite{zuckerkandl1962}. Similarly, although to a lesser degree, progress has been made on methods for efficient exploration of the space of tree topologies\parencite{stamatakis2014}, which is fundamentally challenging due to its combinatorial complexity. Indeed, for $n$ taxa, there are $(2n-3)!!$ possible rooted tree arrangements, where $n!!$ denotes the semifactorial of $n$ --- even a small dataset of ten taxa can be enumerated by 34 million unique rooted trees. Moreover, finding the global optimal tree is NP-hard for all major optimality criteria (e.g., maximum parsimony~\parencite{foulds1982}, minimum evolution~\parencite{day1986}, maximum likelihood~\parencite{roch2006}). Methods such as linear programming~\parencite{catanzaro2012} or branch and bound~\parencite{Hendy1982-dy} can provide exact solutions, but are practically limited to problems with $\approx$ 15 or fewer taxa. To overcome these challenges, the overwhelming majority of state-of-the-art software (e.g., \texttt{MrBayes}~\parencite{huelsenbeck2001}, \texttt{PAUP}~\parencite{Wilgenbusch2003-mu}, \texttt{BEAST}~\parencite{Drummond2007-za}, \texttt{PAML}~\parencite{Yang2007-ez}, \texttt{RAxML}(\texttt{-NG})~\parencite{stamatakis2014, kozlov2019}, \texttt{FastME}~\parencite{lefort2015}, \texttt{IQ-TREE}~\parencite{nguyen2015, minh2020}) rely on hand-engineered search heuristics to perform tree topology optimisation or Bayesian analysis. These are traditionally based on subtree pruning and regrafting (SPR) and tree bisection and reconnection (TBR) operations, which have empirically been shown to be the best available methods for exploring tree topology space~\parencite{stamatakis2014, Park2010-zq}.

However, such methods still have limitations. First, hill climbing using heuristic approaches necessitates multiple evaluations of the objective function to pick the best move. While these heuristics are still polynomial, exhaustive exploration of single SPR operations is quadratic in complexity, and paired operations (two sequential SPR changes) are quartic. Second, all the aforementioned tree arrangements are prone to being trapped in local optima and even if a global optimum is found, terraces of trees with identical quality exist~\parencite{sanderson2011}. The challenge of exploring tree space is exacerbated when concatenating multiple genes in supermatrices~\parencite{rokas2003, dequeiroz2007, Chernomor2016-vi} or when using genomic-scale datasets which require extensive computational resources.

To address these shortcomings, we propose \textsb{GradME}, a new direction for tree topology inference which expands the problem space using a continuous rather than discrete parameterisation of a phylogenetic tree. Generally, aside from considering metrics (e.g., distances in tree space)~\parencite{St_John2017-vq,Dinh2017-og,Billera2001-il,Chernomor2016-vi}, performing topological search in a continuous tree space has rarely been explored (for recent work in hyperbolic spaces, see~\parencite{matsumoto2021, wilson2021, macaulay2023, mimori2023}). Furthermore, very few approaches have made use of gradient-based tree proposals~\parencite{matsumoto2021, zhang2018, nesterenko2022,Dinh2017-og}. Although maximum likelihood and Bayesian inference criteria are more popular and generally considered state-of-the-art~\parencite{huelsenbeck2001, stamatakis2014, Drummond2007-za, minh2020, Wilgenbusch2003-mu, ayres2012}, the GradME framework optimises tree topology under a balanced minimum evolution (BME) criterion~\parencite{Pauplin2000-vk,Desper2002-uy} using distance matrices as an input. This criterion is well-principled~\parencite{Kidd1971-sk} but generally performs worse than likelihood-based~\parencite{price2009, yang2012, lefort2015}. However, the framing of the minimum evolution criterion~\parencite{Kidd1971-sk, Rzhetsky1992-qj} has been proven to be statistically consistent~\parencite{Desper2004-ij,felsenstein2004} and has repeatedly shown good (although not state-of-the-art) performance in various settings~\parencite{Kuhner1994-cf, kumar2000, Gascuel2006-sr, lefort2015}. 

To better explore the space of possible trees, we expand the space over which we need to search. Our novel vector representation of a phylogenetic tree, Phylo2Vec~\parencite{penn2023}, has a natural continuous extension, allowing us to improve the ability to search parts of this space. Appealing to a common analogy that casts the optimal tree search problem as finding a needle in a haystack, our approach observes a much bigger haystack, but the hay is in very large bundles, many of whom have a needle, and for these bundles we have access to a (weak) magnet. Providing details to this analogy, the size of the usual phylogenetic haystack with $n$ taxa is $(2n-3)!!$~\parencite{Felsenstein1978-qc}, while we search a much larger haystack of size $(n!)^2$. There are $n!$ bundles in this larger haystack, each of which contains $n!$ trees, but for any tree, $2^{n-1}$ bundles will contain that tree. Although the proportion of bundles containing a needle shrinks exponentially, we propose a novel approach (Queue Shuffle) that chooses bundles that should be closer to one with a needle. For any given bundle, we also introduce a continuous objective function that can be efficiently traversed using gradient descent approaches (the weak magnet) developed for large-scale machine learning problems~\parencite{Kingma2014-ow, loshchilov2018, shazeer2018}. This continuous objective facilitates enormous changes to tree topology in a single step in a direction that improves the objective function. After searching any given bundle using the continuous objective, we use Queue Shuffle, which improves the switch towards the next bundle to search. This counterintuitive approach offers a new addition to the existing heuristic methods used for topological inference, outperforming the current state-of-the-art but as currently stands with a larger overall complexity comparable to other minimum evolution approaches such as Neighbour Joining.

\section*{Results}
\subsubsection*{Tree traversal in continuous space} 
\begin{figure*}[htbp]
    \centering
    \captionsetup[subfigure]{justification=raggedright,singlelinecheck=false}
    \begin{subfigure}{\textwidth}
        \caption{\textsb{Dataset: Primates}}
        \centering \includegraphics[width=0.9\textwidth]{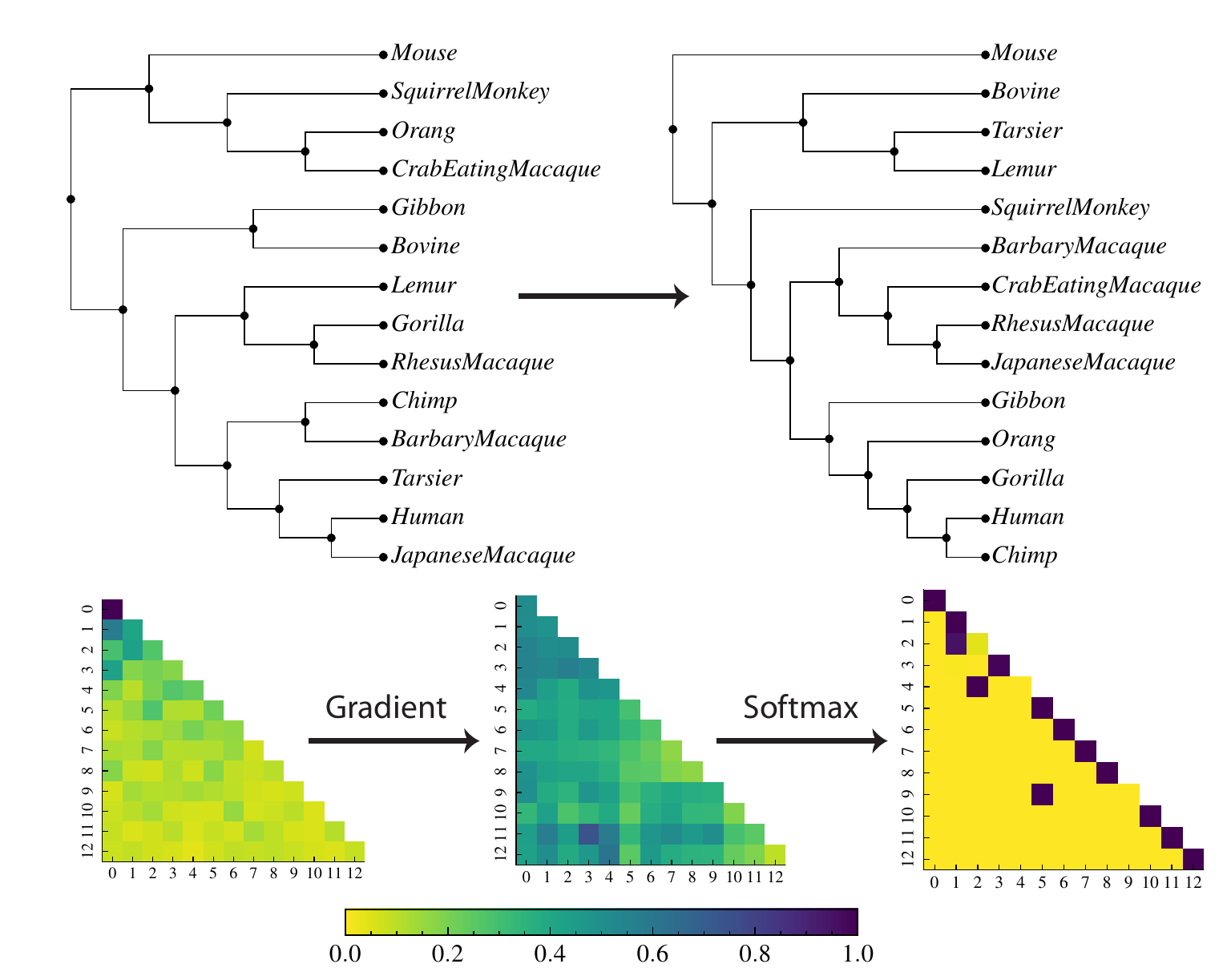}
        \label{fig:big_jump}
    \end{subfigure}
    \begin{subfigure}[t]{0.45\textwidth}
         \centering
         \caption{\textsb{Dataset: Simulated (20 taxa, 100k sites)}}
         \vspace{2em}
         \includegraphics[height=6.25cm]{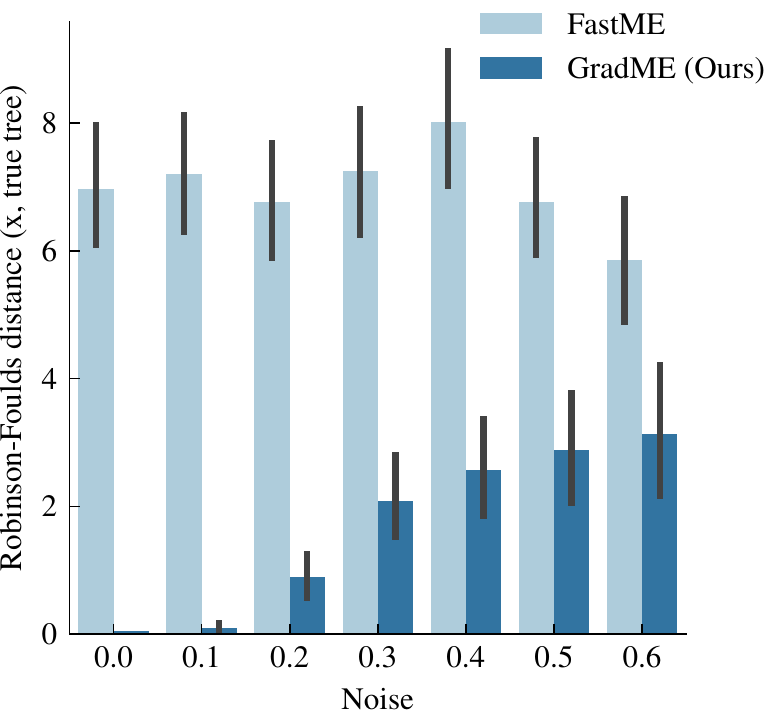}
         \label{fig:sim_rooting}
     \end{subfigure}
    \begin{subfigure}[t]{0.45\textwidth}
         \centering
         \caption{\textsb{Dataset: Jawed}}
         \vspace{2em}
         \includegraphics[height=6.25cm]{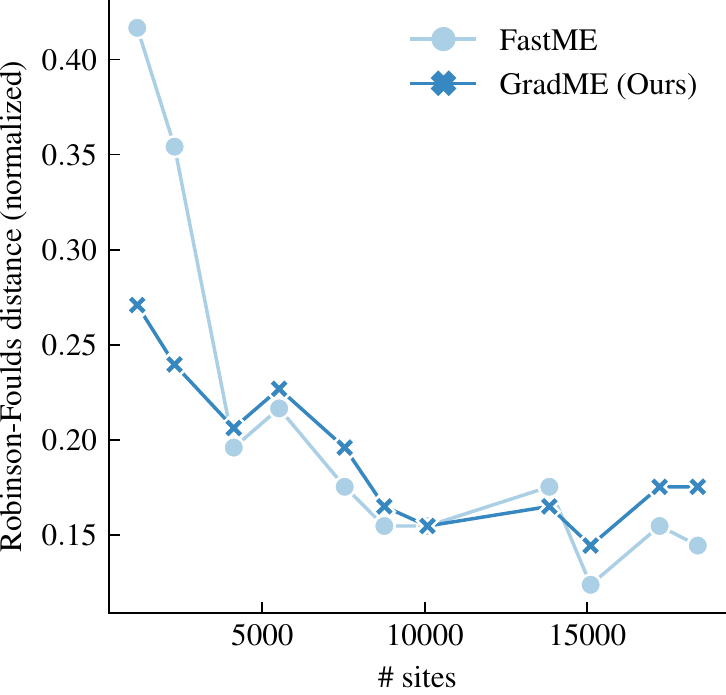}
         \label{fig:jawed_verts}
    \end{subfigure}
    \caption{Results on empirical data \textsb{(a)} Starting from a random tree, represented by an $n\times n$ stochastic matrix, we compute the continuous gradient, apply softmax activation and increment the original matrix. In a single step, our gradient finds the correct tree at a distance of 6 subtree-prune and regraft moves from the random starting tree. \textsb{(b)} Simulating ultrametric trees of 20 taxa and 100,000 sites under an LG model of protein evolution. We add random uniform noise to all branch lengths to simulate departures from ultrametricity. Compared to the true tree via Robinson-Foulds distance, light blue bars are midpoint rooting the best \texttt{FastME} tree and dark blue bars are the inferred root from our approach. \textsb{(c)} Phylogenies for jawed vertebrates, where the number of genes (hence sites) are reduced to be more clocklike. Normalised Robinson-Foulds distance are shown between the best \texttt{ASTRAL}~\parencite{Zhang2018-iw} tree, the best unrooted \texttt{FastME} tree which has been midpoint rooted (light blue) and our inferred rooting algorithm (dark blue). Performance for \texttt{FastME} reduces when the number of sites is small.}
    \label{fig:results}
\end{figure*}

For any choice of label ordering, our approach admits a continuous gradient across $n!$ trees for $n$ leaves. This gradient, which can be obtained readily via automatic differentiation, can rapidly traverse tree space to find trees with a close to optimal objective value. Fig.~\ref{fig:big_jump} shows a single gradient step for the small Primates dataset~\parencite{paradis2004,Yang2007-ez}. Simply subtracting the gradient from a random initial tree, followed by softmax activation, results in an almost discrete $W$ which corresponds to the best BME tree for a given substitution model. Note that if more gradient steps were taken, the $W$ matrix would quickly become discrete (from Lemma~\ref{lem:discrete}). The jump taken corresponds to six subtree-prune and regraft moves~\parencite{paradis2004}.

For larger alignments such as the popular Eutherian dataset~\parencite{Song2012-ql}, a single gradient step can result in 14 to 18 SPR moves. While the number of SPR moves achieved is large, this is achieved with a substantial increase in overall computational complexity when compared to FastME. We note that the gradient step size is dependent on the data and, as expected, greatly reduces as we approach an optimum.

\begin{table*}[htbp]
\centering
\caption{Balanced minimum evolution loss scores for 11 phylogenetic benchmark datasets. Lower is better. Scores from BioNJ and FastME were obtained following the implementations in \texttt{ape}~\parencite{paradis2004} using the same distance matrix as GradME. The distance matrix was estimated from a GTR+$\Gamma$ model via maximum likelihood~\parencite{yang2006}. Our GradME approach always starts from a uniform tree distribution (every tree is equiprobable) with a random taxon ordering (optimised by Queue Shuffle). The best performing approaches for each dataset are denoted in bold. GradME either equalled or performed better than FastME. The topological accuracy, measured as one minus the Robinsons-Foulds distance is shown between GradME and FastME and GradME and a maximum likelihood gold standard from IQ-TREE also using a GTR+$\Gamma$ model.}
\begin{tabular}{
    l *3{r}%
    >{\raggedleft\arraybackslash}p{3cm}%
    >{\raggedleft\arraybackslash}p{4cm}
}
\toprule
Dataset & BioNJ & FastME             & GradME             & Topological accuracy between GradME and FastME & Topological accuracy between IQ-TREE and the best distance tree \\
\midrule
DS1     & 0.3118613         & \textbf{0.3101232} & \textbf{0.3101232} & 1.00 & 0.54 \\
DS2     & 3.725205          & \textbf{3.7239944} & \textbf{3.7239944} & 1.00 & 0.77 \\
DS3     & 8.0115913         & \textbf{8.0075588} & \textbf{8.0075588} & 1.00 & 0.97 \\
DS4     & 2.2528503         & \textbf{2.2447615} & \textbf{2.2447615} & 1.00 & 0.68 \\
DS5     & 6.3077156         & \textbf{6.2606057} & \textbf{6.2606057} & 1.00 & 0.70 \\
DS6     & 0.6249236         & 0.6228563          & \textbf{0.6219367} & 0.87 & 0.67 \\
DS7     & 9.9174641         & \textbf{9.882138}  & \textbf{9.882138}  & 1.00 & 0.91\\
DS8     & 1.337924          & \textbf{1.3252984} & \textbf{1.3252984} & 1.00 & 0.82 \\
DS9     & 0.3788481         & \textbf{0.3788481} & \textbf{0.3788481} & 1.00 & 0.66 \\
DS10    & 1.1286037         & \textbf{1.1247627} & \textbf{1.1247627} & 1.00 & 0.78 \\
DS11    & 1.313921          & 1.3096422          & \textbf{1.3096415} & 0.88   & 0.53 \\
\bottomrule
\end{tabular}
\label{tab:bme_results}
\end{table*}

\subsubsection*{A comparison to benchmark phylogenetic data sets}
Table~\ref{tab:bme_results} presents a comparison of GradME with neighbour joining (BioNJ) and FastME (subtree-prune and regraft version) over 11 popular phylogenetic benchmark datasets~\parencite{Whidden2015-jn}. Both neighbour joining and FastME are only able to infer a minimum length unrooted tree, and therefore we compare estimates only on unrooted trees. We always initialise our algorithm with a uniform, equiprobable tree, where the starting taxon labelling is random and optimised using Queue Shuffle. We estimate tree using distances from a GTR+$\Gamma$ model estimated via maximum likelihood (see Appendix~\ref{append:est} for details). As expected, FastME consistently outperforms BioNJ, with lower BME loss on all alignments. On the other hand, GradME always achieves a better or equal loss compared to FastME. We observe similar results when using different substitution models (e.g., F81). In the two examples where GradME does better than FastME, the topological accuracy, measured by one minus the Robinson-Foulds distance~\parencite{robinson1981} is close to 0.9, suggesting FastME has converged to a similar tree. We note that FastME's performance is generally worse when using the nearest neighbour interchange heuristic (instead of the SPR-based heuristic). When compared to a maximum likelihood gold standard (IQ-TREE~\parencite{minh2020}), the best distance method does not recover the same tree as that from maximum likelihood, but in some cases, is very close (e.g., DS3 and DS7). Finally, we note that while GradME outperforms FastME, it is orders of magnitude slower and in most of the data sets FastME finds the same optimal tree as GradME. 

\subsubsection*{Rooting ultrametric trees}
Despite being applicable to the unrooted problem, our approach, at its core, works with rooted trees. As previously discussed, if we assume the existence of a distant outgroup, then the balanced minimum evolution objective can be used to optimise a rooted phylogenetic tree. In Appendix~\ref{sec:bme_rooted}, we show that, given an ultrametric unrooted tree, the optimal rooting maximises a heuristic for the root-to-tip distance in the tree. Equivalently, the optimal rooting ensures that the root is estimated to be the maximal possible distance back in time. This is not an immediately biologically plausible objective for the root. Indeed, the cornerstone of balanced minimum evolution is finding the tree of minimum length, and it hence seems counter-intuitive to require the root that is the maximum distance backwards in time (though this does in fact minimise the tree length). However, our assumption of a distant ancestor means that the root of our tree must be the point that is furthest backwards in time. In particular, this means that the evolutionary direction needs to be away from the root. By setting our root such that the root-to-tip distance is maximised, we ensure that the root satisfies this constraint. However, this property does not hold for trees that are not ultrametric - in these cases, the root will be drawn towards branches with higher mutation rates.

While this property only holds for ultrametric trees, our approach still works well for near clock-like trees. As an experiment, we draw small (20 taxa) random ultrametric phylogenies with a total length of one, and simulate 100,000-residue protein sequences~\parencite{paradis2004} down these trees under an LG~\parencite{Lee2015-es} model of protein evolution, assuming random uniform amino acid base frequencies. In the ultrametric cases, all taxa are equidistant to the root, which corresponds to a strict molecular clock. We add uniform noise to all branch lengths to simulate departure from a strict clock. Fig.~\ref{fig:sim_rooting} shows the Robinson-Foulds~\parencite{robinson1981} distance from the true tree to the \emph{midpoint-rooted} best \emph{unrooted} \texttt{FastME} tree (when SPR moves were used by \texttt{FastME}), and the distance to our inferred rooted tree. We see that when the tree is ultrametric, or close to ultrametric, our approach recovers the correct rooted tree. As expected, an increase in noise leads to a decrease in topological accuracy, although our approach still performs substantially better than midpoint rooting. We note that uniform noise is unlikely to be biologically realistic. Instead, deviations from a strict clock are more likely to be heterogeneous in certain clades or internal branches. However, for small departures, we believe our algorithm to reliably infer the correct tree and root simultaneously. 

We implement our rooting algorithm on the popular mammal data from~\parencite{Song2012-ql}. We infer a rooted tree via Queue Shuffle and also midpoint root the best FastME tree. Both trees, unrooted, have the same balanced minimum evolution loss, but our rooted loss is less than the FastME midpoint rooted loss. Our rooted tree correctly identifies \textit{Gallus gallus} (red junglefowl) as the outgroup, while midpoint rooting pairs \textit{Gallus gallus} with \textit{Ornithorhynchus Anatinus} (platypus) (see Fig.~\ref{fig:Songtree} for the rooted phylogenies).

\begin{figure*}[htbp]
    \centering \includegraphics[trim={0.5cm 0.5cm 0.5cm 0.5cm},clip, width=\textwidth]{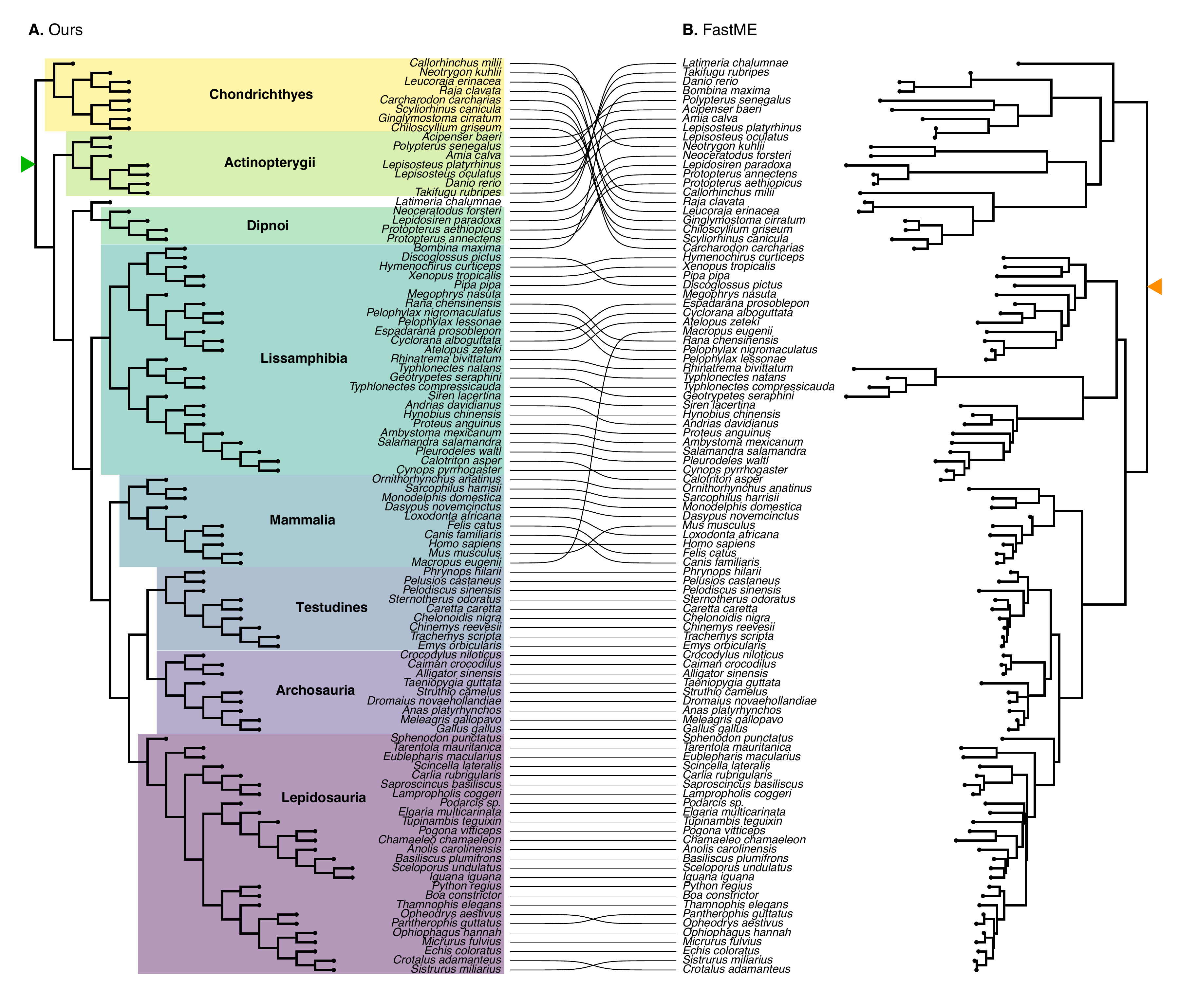}
    \caption{Phylogenetic inferences of the jawed vertebrates' phylogeny using the two most ultrametric loci from a data set of 99 taxa and 4593 genes~\parencite{irisarri2017}. \textsb{(a)} Inference using our approach leads to high accuracy in identifying the root and all major jawed vertebrate taxa. Note that, we do not estimate branch lengths, but only topology via balanced minimum evolution \textsb{(b)} inference using FastME and midpoint rooting leads to widespread error, primarily and critically near the root of the process.}
    \label{fig:jverts_tree}
\end{figure*}

\subsubsection*{Rooting the phylogeny of all jawed vertebrates}

To perform a more detailed evaluation of our framework, we tested GradME's robustness for topological inference by finding the root of the large jawed vertebrates dataset from~\parencite{irisarri2017} with 99 taxa and 4593 genes. Given the reliance of our method on ultrametric data for inference of the root, we first made a fast measure of the ultrametricity of each gene-tree. To do this, we inferred the phylogeny of each gene using GradME, followed by midpoint rooting. The coefficient of variation in root-to-tip lengths was taken as a measure of ultrametricity. We then concatenated ranked genes into supermatrices including decreasing numbers of genes, and examined the performance of GradME with midpoint rooting against our method. All inferences were performed using the LG amino-acid substitution model to maintain simplicity. We placed special focus on our ability to use small portions of data for recovering the main groupings of vertebrates; these key groupings include the root separating cartilaginous (Chondrichthyes) versus boned vertebrates, ray-finned fishes (Actinopterygii), and the major groups of tetrapods and amniotes (amphibians, mammals, archosaurs, turtles, and lizards and relatives).

A small numbers of genes with an ultrametric signal were generally sufficient for resolving many of the major lineages of vertebrates using both midpoint rooting and our approach (Fig.~\ref{fig:jawed_verts}). For larger numbers of genes, midpoint rooting and our approach are broadly similar. However, at the smallest numbers of genes (0.05\%, 2 genes), midpoint rooting was unable to recover many of the early relationships among vertebrates, such as the root, monophyly of cartilaginous fishes, ray-finned fishes, Tetrapoda, or the mammals. Even small amounts of data (1460 amino-acids of 1,964,439; 0.07\% of the original data) were sufficient for GradME to resolve the root as well as every major grouping of jawed vertebrates accurately (Fig.~\ref{fig:jverts_tree}). The only exception was the controversial position of the Coelacanth, which was found to be sister of Dipnoi (lungfish) rather than the more widely accepted position as sister of Dipnoi plus Tetrapoda. While this remarkable performance under the simple LG model is in part attributed to the informative nature of highly ultrametric genes, our tree topology demonstrates the superiority of our approach in accuracy and efficiency over other fast methods in phylogenetics.

\section*{Discussion}
We have introduced a new approach for exploring the vastness of tree space. Counterintuitively, our approach explores a much bigger space than the space of possible trees, but this larger space allows for new ways to find the best tree. The key to our method's success lies in transforming the phylogenetic tree search problem from a discrete to a continuous one, allowing us to achieve superior performance. To our knowledge, this is the first time a continuous, differentiable objective function for the inference of tree topology has been proposed, and it opens new possibilities for phylogenetic inference. Bayesian phylogenetics can be regarded as the most robust framework for inferring phylogenies, but has been to-date limited by the poor ability of random walk Metropolis-Hastings algorithms to explore tree space~\parencite{Betancourt2017-fl}. More efficient Hamiltonian Monte Carlo samplers have been proposed~\parencite{Dinh2017-og} to tackle this problem, and our framework presents a new avenue to jointly explore topology and branch lengths with efficient samplers. A remaining limitation of our approach is the need to shuffle labels to fully explore the space of all possible trees, and while the approach we use, Queue Shuffle, is mathematically and practically powerful, this step is still discrete. The possibility of permutation distributions such as the Gumbel-Sinkhorn distribution could allow for a fully differentiable algorithm. Finally, the complexity of our approach is $\mathcal{O}(n^5)$, which easily allows for large phylogenies up to a thousand, but not tens of thousands. However, computation on GPUs or TPUs in parallel can facilitate computational tractability. 

A major benefit of our approach is that it naturally enables the estimation of the root node, which has been a long matter of interest in the biological sciences~\parencite{Huelsenbeck2002-hb,Tria2017-cw,Naser-Khdour2022-qz}. For genes where a strict clock is a reasonable assumption, our method of traversing tree space in large steps reliably estimates both the correct tree topology and the root. Our approach will likely be useful in settings where genetic sequences are contemporaneous and time for measurable evolution is short, such as early epidemics or nosocomial settings. However, as we showed analytically, our approach will have reduced performance when considering rate heterogeneity and departures from a strict clock. 

Tests on the relationships among jawed vertebrates demonstrate that even minimal amounts of data can be sufficient for our method to reach high accuracy in topology and root estimates. These results are consistent with previous work on large amounts of genome-scale data showing that clocklike loci to be the most suitable for phylogenetic inference~\parencite{Vankan2022-xz}. Furthermore, our approach is effective with negligible amounts of data – where other methods are ineffective – making it a powerful addition to the existing toolkit for addressing recalcitrant questions of the tree of life.

Our approach is based on the minimum evolution principle, which has repeatedly shown to produce fast and accurate inference. Nonetheless, an interesting area for further study is to extend the continuous path length formulation to approximations of traditional phylogenetic likelihoods~\parencite{felsenstein1983}. This would be particularly beneficial for implementation in Bayesian inference, since tree topology inference is a major obstacle to large hierarchical models~\parencite{Suchard2003-mh,Duchene2016-mh}. Our method is therefore a step towards more efficient sampling of the complex posterior distributions over tree topology. 

\section*{Methods}
In the following, we describe GradME, a distance-based method for continuous phylogenetic inference of rooted and unrooted trees using gradient descent. The framework can be divided into three components: i) a continuous tree representation based on Phylo2Vec~\parencite{penn2023}, a bijective integer representation of phylogenetic trees; ii) gradient-based optimisation using a continuous version of the balanced minimum evolution criterion~\parencite{Pauplin2000-vk, Desper2002-uy}, iii) Queue Shuffle, a method to shuffle the integer-to-taxon mapping underlying Phylo2Vec for full tree space exploration. The overall approach works for both rooted and unrooted trees.


\subsubsection*{Balanced minimum evolution}~Popular objective functions to infer the optimal tree from phylogenetic data include maximum parsimony~\parencite{Fitch1967-dr}, maximum likelihood~\parencite{felsenstein1983} and minimum evolution~\parencite{Saitou1987-wg}. Maximum likelihood and minimum evolution are provably statistically consistent~\parencite{Desper2004-ij,felsenstein2004}, whereas maximum parsimony can be inconsistent under certain conditions~\parencite{Felsenstein1978-jt}. For small to moderate sized phylogenies, methods based on maximum likelihood (and Bayesian extensions) are generally considered state-of-the-art~\parencite{stamatakis2014, Drummond2007-za, minh2020, Wilgenbusch2003-mu}. However, approaches based on minimum evolution (ME) have also shown to yield adequate performance~\parencite{Kuhner1994-cf, Gascuel2006-sr, lefort2015, kumar2000, roch2006}. The first introductions of the minimum evolution (ME) paradigm~\parencite{Kidd1971-sk, Rzhetsky1992-qj} sought to express evolutionary relationships through dissimilarity. They proved that, given unbiased estimates of the true evolutionary distances, the true phylogeny has an expected length shorter than any other possible phylogeny – thereby establishing the principled ME criterion. Currently, the best performing ME approach is that of balanced minimum evolution (BME)~\parencite{Pauplin2000-vk,Desper2002-uy}, with \texttt{FastME}~\parencite{lefort2015} being a popular software implementation. Its objective function can be written as:
\begin{equation}
    \mathcal{L}(T) = \sum_{i,j}D_{ij}2^{-e_{ij}}
    \label{eq:bme_objective}
\end{equation}
where $D_{ij}$ denotes a distance (e.g., based on molecular sequence data) between two taxa $i$ and $j$ and $e_{ij}$ the number of branches in the path between taxa $i$ and $j$ (the path length~\parencite{Semple2004-my}).
This objective can be computed in a numerically stable fashion using the log-sum-exp trick (see Appendix~\ref{append:code} for an example snippet). A widely used approach to estimate the optimal tree greedily~\parencite{Desper2004-ij, Gascuel2006-sr} is the neighbour joining (NJ) method~\parencite{Saitou1987-wg}. When neighbour joining is based on an additive distance measure, it reconstructs a unique tree, but still performs well with near-additive trees~\parencite{Atteson1999-jd} and under small perturbations in the data~\parencite{Mihaescu2009-es}. However, despite these highly favourable properties, further heuristic optimisation on a NJ tree using SPR moves have proven to be even more accurate~\parencite{lefort2015}. Once a tree topology is found, quadratic algorithms exist for estimating the branch lengths~\parencite{Mihaescu2008-bo} as well as efficient approaches for molecular clock dating~\parencite{To2016-vd}.

\subsubsection*{Balanced minimum evolution for rooted trees}

Inference using BME is always restricted to unrooted trees~\parencite{Semple2004-my, catanzaro2022} with rooting chosen after inference through heuristics (e.g., midpoint rooting) or via a molecular clock (e.g., for serially sampled data). However, it is often of interest to find the optimal rooted tree for a set of taxa, as this provides extra biological context (e.g., to represent evolutionary paths).

In an unrooted tree, the BME objective function (Eq.~\ref{eq:bme_objective}) provides an efficient way of calculating the total length of a tree where the branch lengths are the least squares estimators for approximating each $D_{ij}$ with the distance from nodes $i$ to $j$ in the tree. However, this result does not hold in a rooted tree, as the addition of a root changes many of the path lengths. To remedy this, we consider adding a “root taxon” to the tree by joining it to the root node as taxon $n$. If the tree is roughly ultrametric, then we expect
\begin{equation}
    D_{ni} \approx D^*  \quad \forall i \neq n
\end{equation}
where $D^*$ is the (assumed constant) root-to-taxa distance. Of course, we do not know the sequence of the root but, as we will show, the value of $D^*$ is unimportant — it is instead simply important that it is independent of $i$. Adding this root taxon as a leaf node transforms the tree from being rooted to being unrooted, where standard BME can be used. From this assumption, we prove two lemmas to ensure the framework's validity, showing that the optimal unrooted tree is obtained when the variation in the root-to-taxa distance is sufficiently small (Lemma~\ref{lem:rooted1}), and subsequently that, in all cases, the optimal rooting for an unrooted tree solves a biologically plausible optimisation problem (Lemma~\ref{lem:root_min}).

First, Lemma~\ref{lem:rooted1} shows that, if
\begin{equation}
    |D_{ni} - D^*| < \delta \quad \forall i \neq n
\end{equation}
then, using $e_{ij}^u$ and $e_{ij}^r$ to denote path lengths in the unrooted tree ($u$) containing taxon $n$ and the rooted tree ($r$) formed by removing taxon $n$,
\begin{equation}
     \bigg|\sum_{i=0}^n\sum_{j=0}^nD_{ij}2^{-e^u_{ij}} - \sum_{i=0}^{n-1}\sum_{j=1}^{n-1}D_{ij}2^{-e^r_{ij}} - D^*\bigg| \leq \delta
\end{equation}
where $\delta$ denotes a small number. Hence, the difference between the rooted and unrooted objective functions is approximately equal to the constant, $D^*$. Thus, for sufficiently small $\delta$, by the discreteness of tree space, we can see that, if it is unique, the optimal unrooted tree under the \textit{rooted} objective (using $e_{ij}^r$) will be the same as that under the unrooted objective (using $e_{ij}^u$), when the root taxon is used as an additional leaf.

Subsequently, Lemma~\ref{lem:root_min} shows that, for the correct unrooted tree, the BME-optimal rooting maximizes a simple heuristic (defined in Definition~\ref{def:distance}) for the root-to-tip distance. Equivalently, the optimal rooting ensures that the root is estimated to be the maximal possible distance back in time.

This is not an immediately biologically plausible objective for the root. Indeed, the cornerstone of BME is that we want the tree of \textit{minimum} length, and it hence seems counter-intuitive to require the root that is the \textit{maximum} distance backwards in time (though, by Lemma~\ref{lem:root_min}, this does create the minimum length tree). However, the root of our tree must be the point that is furthest backwards in time. In particular, this means that the evolutionary direction needs to be away from the root. By setting our root such that the root-to-tip distance is maximised, we ensure that the root satisfies this constraint.

This method shares a similar motivation with midpoint rooting, which also seeks to maximise a heuristic for root-to-tip distance. However, the heuristic used in midpoint rooting uses only the two taxa which are furthest apart, while our rooting method uses distances from all taxa. Thus, we expect our method to be more robust, particularly as large inter-taxa distances are difficult to estimate, meaning that the additional information used in our heuristic should help to reduce errors. This is evidenced in Fig.~\ref{fig:Songtree}, where midpoint rooting leads to incorrect root placement.

Nonetheless, this property will not hold if the tree is not ultrametric. If taxa evolve at different rates at different times throughout the tree, then the root will be drawn towards taxa with high evolutionary rates. Thus, caution must be used when applying our rooted algorithm to such trees, although the unrooted algorithm will still give a correct unrooted tree. In this case, it may be best to find the optimal unrooted tree topology and then solve the rooting problem for this tree, rather than finding the optimal rooted tree, as the former will reduce the skewing effect of the heterogeneity in evolutionary rates. Because of this, the algorithm introduced in this paper has the flexibility to find the optimal unrooted or rooted tree.

\begin{figure*}[htbp]
    \centering
    \includegraphics[width=0.72 \textwidth]{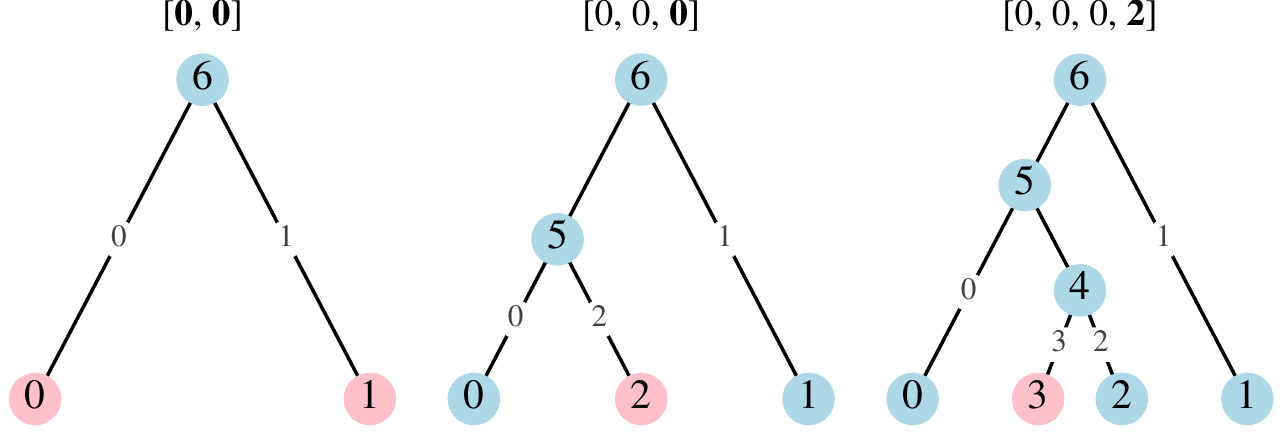} 
    \caption{An example of the left-to-right construction of the ordered tree $v = [0,0,0,2]$.}
    \label{fig:ltr}
\end{figure*}

\subsubsection*{An ordered bijection to tree space}~Previously, we introduced Phylo2Vec~\parencite{penn2023}, a novel bijection between the space of phylogenetic trees and a space of integer vectors. In contrast to other bijections such as permutation matchings~\parencite{diaconis1998}, changes in Phylo2Vec correspond to smooth changes in the tree space, e.g., single changes in a Phylo2Vec vector correspond to a limited set of SPR changes. On the other hand, Prüfer codes~\parencite{C_Chen2000-ak} form a bijection to the space of all $m$-ary trees, meaning that there is no guarantee to sample binary trees from random Prüfer sequences.

Here, we focus on the notion of \emph{ordered trees} from~\parencite{penn2023}, where it is possible to construct a tree from its vector in linear time. An ordered tree can be thought of as a birth process, such that when a birth occurs, the original node continues to live and retains its label, while the new node receives an incremented label. Accordingly, we introduce an equivalent but more intuitive tree construction process for these ordered trees (see Fig.~\ref{fig:ltr} for an example). We begin with two leaf nodes and two edges labelled 0 and 1. We then append nodes by joining them, \emph{in order}, to edges connecting leaf nodes to the tree. This tree construction process can be summarised by a single vector $\boldsymbol{v}$, with $v_0 = v_1 = 0$ and, for $m \geq 2$, $v_m$ being the label of the edge to which node $m$ is appended. Each index in $v_m$ is subject to the simple constraint:

\vspace{-1cm}
\begin{align}
\label{eq:disc}
    & v_0 = v_1 = 0 \quad \text{and} \\ 
    & v_m \in \{0,1,...,m-1\} \quad \forall \quad m \geq 2 \nonumber
\end{align}
\vspace{-0.5cm}

which is equivalent to the definition of ordered trees in~\parencite{penn2023}. Intuitively, a tree is ordered if, starting with a branch connecting the root to taxon 0, the taxa can be added in order of their label by appending a new branch to a \textit{terminal} branch of the existing tree (i.e. a branch connecting a leaf node to the rest of the tree). Thus, the ordering of the taxa is in some sense ``natural'' for each possible ordered tree.  It is proved in Lemma~\ref{lem:ltr} that for ordered $\boldsymbol{v}$, the algorithm presented above and the more general Phylo2Vec algorithm in~\parencite{penn2023} produce equivalent trees.

Note that, for a fixed integer-taxon labelling, the number of ordered trees is a subset of the number of possible trees. We discuss an efficient method to remedy this problem and explore all tree space called the Queue Shuffle.

\subsubsection*{A continuous representation of a tree}
We introduce continuous, probabilistic, representation of trees using a square matrix $W$ which gives the distribution of a random ordered vector $\boldsymbol{v}$ with independent entries such that $W_{ij} = \mathbb{P}(v_i = j)$. Given Eq.~\ref{eq:disc}, $W$ is a lower-triangular, stochastic matrix (row sums to 1). Thus, $W$ can probabilistically represent any \emph{ordered} phylogenetic tree (a space of $(n-1)!$ trees). A simple approach to determining the most likely single tree from $W$ is to take the column-wise argmax, yielding a single tree $\boldsymbol{v}$.

\subsubsection*{Gradient-based optimisation using the BME criterion}
Using this continuous representation, we can find the optimal ordered tree. Defining $f(\boldsymbol{v})$ to be the BME objective function for the tree generated by an ordered vector $\boldsymbol{v}$, we then create a continuous objective, $F(W)$ by $    F(W) = \mathbb{E}\big[f(\boldsymbol{v})\big]$.

The calculation of $F(W)$ follows our new method of constructing ordered trees from the vector $\boldsymbol{v}$. For a fixed (randomly chosen) tree, we define $e_{ij}^k$ to be the path length between nodes $i$ and $j$ when nodes $0,1,...,k-1$ have been added to the tree (for $i,j < k$). Note that to find the rooted objective function, we initialise with $e_{10}^2 = e_{01}^2 = 2$, while to find the unrooted objective, we initialise with $e_{10}^2 = e_{01}^2 = 1$ (while the Phylo2Vec representation is an inherently rooted representation, this unrooted objective finds the tree length if the root were removed from the random rooted tree with distribution given by $W$). This is because, in a tree where the only leaf nodes are $0$ and $1$, these nodes are a path length of 2 apart if there is also a root (as the root is on this path) while otherwise, they are a path length of 1 apart.

If node $k$ is appended to the edge joining either node $i$ or $j$ to the tree, then $e_{ij}^{k+1} = e_{ij}^k + 1$; otherwise, $e_{ij}^{k+1} = e_{ij}^k$. Similarly, if node $k$ is appended to the edge joining node $i$ to the tree, then $e_{ik}^{k+1} = 2$ and otherwise, if node $k$ is appended to the edge joining node $x$ to the tree, then $e_{ik}^{k+1} = e_{ix}^k + 1$. Thus, using $V_k$ to denote the random value of $v_k$, we can write

\begin{align}
    &e^{k+1}_{ij} = e^k_{ij} + G^k_{ij}(V_k) \\
    &\Rightarrow e^{n-1}_{ij} = \sum_{k=2}^{n-2}G^k_{ij}(V_k) + e^2_{ij}
\end{align}
for some functions $G^k_{ij}$ which are derived explicitly in Lemma~\ref{lem:cont_obj}. Importantly, each term in the sum is independent, and hence, in Lemma \ref{lem:cont_obj}, a closed iterative system for the quantities $E_{ij}^{k} = \mathbb{E}(2^{-e_{ij}^k})$ can be calculated for $i<j$ as
\begin{equation}
        E_{ij}^{k+1} = \left\{ \begin{matrix}
            E^{k}_{ij} \bigg[1 - \frac{1}{2}(W_{ki} + W_{kj})\bigg] & \text{if $i < j < k$}\\
              \bigg[\frac{1}{2}\sum_{x\neq i}E_{ix}^kW_{kx}\bigg] + \frac{1}{4}W_{ki} & \text{if $i < k$}
        \end{matrix}\right.
    \end{equation}
    with the remaining values for $i > j$ following by symmetry.

The objective function is a polynomial function of the entries of $W$ and is linear in each fixed entry (that is, the diagonal entries of $\nabla^2 F$ are zero). Thus, by Lemma~\ref{lem:discrete}, there is always a minimum at a “discrete tree'' (that is, at a matrix $W$ where for each row, one value is 1 and all the others are 0). Moreover, this simple form makes it easy to differentiate $F$ analytically, numerically or automatically. Using state-of-the-art automatic differentiation~\parencite{jax2018github}, gradient descent can be used to efficiently minimise $F$ and find the optimal \textit{ordered} tree. 

There may also be minima at non-discrete trees if multiple trees share the same, optimal, objective value. In our rooted optimisation, this is highly unlikely (as two topologically different trees having equal objectives places a dimension 1 condition on the distance matrix $D$, meaning the set of distance matrices for which this happens has measure 0 in the set of possible distance matrices), but when we use our unrooted algorithm, this will occur. This is because, as discussed in \parencite{penn2023}, there are $n-1$ Phylo2Vec vectors which, when the root is removed, give the same unrooted tree. Thus, if multiple rooted vectors giving the same unrooted tree $\mathcal{U}$ are in the same space of ordered trees, then, if $\mathcal{U}$ is the optimal tree under this ordering, the algorithm may converge to a non-discrete $W$. In this case, taking the argmax safely recovers an optimal rooted tree as, by Lemma~\ref{lem:discrete}, all possible trees according to $W$ will have the optimal objective value.

The tree-space induced by our continuous objective function will have local minima whenever changing any single entry of the vector $\boldsymbol{v}$ causes the objective function to increase. As the Hamming distance between vectors $\boldsymbol{u}$ and $\boldsymbol{v}$ is comparable to SPR distance~\parencite{penn2023}, we therefore expect that the discrete subset of our continuous tree-space will be similar in structure to the space induced by SPR moves. However, by starting from a uniformly distributed tree, where all possible ordered trees contribute to the objective, we expect that our algorithm is better able to pick up the ``signal'' from the true optimum, and avoid moving towards suboptimal local minima.

A Python-like algorithm to compute $ \mathbb{E}(2^{-e})$ is shown in Algorithm~\ref{algo:get_edges_exp}. To find the $E^m_{ij}$ terms, there are $\mathcal{O}(m)$ steps of $\mathcal{O}(m)$ (finding the $E_{m-1, j}^m$ terms) and $\mathcal{O}(m^2)$ steps of $\mathcal{O}(1)$ (finding the other $E_{ij}^m$ terms). There are $\mathcal{O}(n)$ values of $m$ that need to be considered, and hence this system can be solved in $\mathcal{O}(n^3)$ time.

\begin{algorithm}
\caption{Compute $E := \mathbb{E}(2^{-e})$ }  \label{algo:get_edges_exp}
\begin{algorithmic}[1]
\Require $n$ \Comment{fixed number of taxa}
\Require $W$ \Comment{stochastic input matrix\parencite{penn2023}}
\State $E = \text{{zeros}}((n, n))$
\For{$i \gets 1$ \text{{ to }} $n - 1$}
    \State $E^* = \text{{zeros}}((n, n))$
    \For{$k \gets 0$ \text{{ to }} $i - 1$}
        \For{$j \gets 0$ \text{{ to }} $k - 1$}
            \State $E^*_{[k,j]} = E_{[k,j]} (1 - \frac{1}{2} (W_{[i-1,j]} + W_{[i-1,k]}))$
        \EndFor
        \State $E^*_{[i,k]}= \sum_{l=0}^{i-1} \frac{1}{2} E_{[l,k]} W_{[i-1,l]} + \frac{1}{4}  W_{[i-1,k]}$
    \EndFor
    \State $E = E^* + E^{*T}$
\EndFor
\State \text{{Return }} $E$
\end{algorithmic}
\end{algorithm}

\subsubsection*{Orderings}
The continuous objective function is only defined for ordered trees, which, for a given labelling of the nodes, is a subset of the whole tree space. Thus, we define the concept of an \textit{ordering} of the nodes, whereby changing the ordering allows the full space of trees to be explored.

\begin{definition}
   \textbf{Ordering:} Suppose that the nodes correspond to taxa with names $N_0,N_1,...N_{n-1}$. We then define an \textit{ordering} of the nodes to be a permutation, $\sigma$ of the set $\{0,...,n-1\}$ such that the node with name $N_i$ is processed as node $\sigma(i)$ by the Phylo2Vec algorithm. It is necessary that the associated Phylo2Vec vector $\boldsymbol{v}(\sigma)$ is ordered.
\end{definition}
Note that we use the phrase “node $x$'' to mean “the node with name $N_x$'' and will always be explicit if we refer to a node by its label.

Given a tree, it is possible to generate a possible ordering of the nodes, $\sigma$, as well as the associated vector $\boldsymbol{v}(\sigma)$. We will do this by labelling the tree as follows, again distinguishing between the leaf node names $N_i$ and their labels (which we will call $l(i)$).

Consider labelling the root node as 0. Then, label the children of this node as $0$ and $1$. One can continue this process inductively, choosing a node, labelled $x$, with unlabelled children and then labelling its children as $x$ and $y$, where $y$ is the smallest unused label. This process terminates when every node has been labelled. As discussed above, suppose that the label of the leaf node with name $N_i$ is $l(i)$. Then, Lemmas \ref{lem:ordering_1}, \ref{lem:ordering_2} and \ref{lem:ordering_3} show that $l$ is a possible ordering of the tree, and provide a method for calculated the associated ordered Phylo2Vec vector $\boldsymbol{v}(l)$.

Given a tree, one can use the previous labelling algorithm to show that there are at least $2^{n-1}$ possible orderings for the tree (as the children of each node can be labelled in either order). The exact number depends on the choice of which internal node is processed at each step (and so, a “balanced'' tree where the leaves have a low generation has more possible orderings than a “ladder'' tree where each leaf has a distinct generation). However, in general, it will be substantially larger than $2^{n-1}$ (we conjecture that for flat trees, it may be factorial in size) and hence there are numerous possibilities which will allow for a global minimum of the objective to be found.
\subsubsection*{Queue Shuffle: changing orderings to explore all the tree space}
\label{subsec:queue_shuffle}
The number of ordered trees (different from ranked trees~\parencite{Collienne2021-eg}), $(n-1)!$, is substantially smaller than the possible number of trees, $(2n-3)!!$ (albeit with a comparable growth pattern in $n$), and hence, while the optimal ordered tree will be closer to the true tree than a very large proportion of trees, it is very unlikely to be exactly equal to the true tree. 

To fully explore tree space, one must shuffle the labels of the leaf nodes in the optimal ordered tree. Simply choosing a uniformly random permutation will lead to extremely inefficient optimisation, as each tree is only possible in approximately $1/2^{n}$ of the possible orderings. Instead, we use the topology of the optimal tree to inform our choice of permutation through a novel approach we call the \textit{Queue Shuffle}. This ensures that the previous optimal tree can be written as an ordered tree in the new ordering, while also ensuring a smooth and efficient path through the space of orderings.

The Queue Shuffle is motivated by the labelling procedure discussed in the previous section, but ensures that the set of internal nodes with a given generation (that is, a given distance from the root) are processed consecutively. That is, we begin by processing all nodes with generation 0 (i.e., the root), then all internal nodes with generation 1, then all internal nodes with generation 2, and continue in this fashion until all internal nodes have been processed.
\\
\\
\noindent
Algorithmically, this can be achieved by a “queue'' of internal nodes to be processed. When an internal node is processed, any of its children that are also internal nodes are added to the back of this queue. Thus, the queue is always in ascending order of generation, and it is simple to show that this ensures that nodes are processed in non-decreasing order of generation. 
\\
\\
\noindent
A crucial feature of this queue is that the child given the same label as its parent is placed \textit{ahead} of the other child in the queue. This ensures that one can, in some way, control the order of processing by choosing the labelling of the children of each node. Moreover, it is vital for the theoretical result presented in the following section. 
\\
\\
\noindent
To add randomness into the labelling procedure, every time an internal node is processed, we randomly choose which child is given the label of their parent, and which child is given the next available label. This provides $2^{n-1}$ possible orderings for each tree. This stochasticity is helpful in ensuring that the algorithm does not get stuck – as discussed in the subsequent section, it ensures that a large class of similar trees will be considered after a few ordering proposals.

An algorithmic description of the Queue Shuffle is provided in Algorithm \ref{algo:qs}.

\begin{algorithm}
\caption{The Queue Shuffle }  \label{algo:qs}
\begin{algorithmic}[1]
\Require $\mathcal{T}$ \Comment{Current Tree}
\Require $\mathcal{N} = \{\nu_0,\nu_1,...\}$ \Comment{set of all non-root nodes}
\State $Q = [\nu_0,\nu_1]$ \Comment{"queue" of nodes to process}
\State $L = \{\nu_0 : 0, \nu_1 : 1\}$ \Comment{node:label mapping}
\State $l_{\text{next}} = 2$\Comment{next available label}
\State $P = []$ \Comment{processed nodes}
\While{$Q \neq []$}
\State $\nu = Q[0]$ \Comment{node to process}
\State Q = Q[1:] \Comment{$\nu$ will be processed}
\State $\text{append}(P,\nu)$ \Comment{$\nu$ will be processed}
\If{$\text{isLeaf}(\nu)$}
\State \textbf{continue} \Comment{move to next node}
\EndIf
\State $a,b = \text{randChildren}(\nu)$ \Comment{get \textbf{randomly ordered} children of $\nu$}
\State $L[a] = L[\nu]$ \Comment{label $a$ with $\nu$'s label}
\State $L[b] =  l_{\text{next}}$ \Comment{give $b$ next available label}
\State  $l_{\text{next}} =  l_{\text{next}} +1$
\State append($Q,a$) \Comment{add $a$ to the queue}
\State append$(Q,b)$ \Comment{add $b$ to the queue}
\EndWhile
\State \text{{Return }} $L$ \Comment{Ordering determined by values of $L$ for leaf nodes}
\end{algorithmic}
\end{algorithm}

\subsubsection*{GradME}
The Queue Shuffle completes our optimisation algorithm. We iteratively find the best ordered tree according to the current ordering and then use Queue Shuffle to change ordering, changing the space of explorable trees. The algorithm terminates when the optimal tree has not been improved upon for a fixed number of iterations (note that, by construction, the previous optimal tree will always be in the new space of ordered trees). In the examples presented in this paper, only tens iterations are needed from some random starting order, and less if a sensible starting ordering (such as from a neighbour joining tree) is used.

We refer the resulting system, combining the continuous tree representation, Queue Shuffle reordering, and the gradient-based optimisation framework using BME, as \textsb{GradME}.

\subsubsection*{Why does Queue Shuffle work?}
A given tree is in the space of ordered trees for at least $2^{n-1}$ orderings. This means that we do not need to find a single optimal ordering, but have exponentially many which will return the true optimal tree. Very loosely considered, being able to explore $n!$ tree space reliably and efficiently with continuous optimisation, Queue Shuffle reduces the inferential task to one that is exponential.

However, while the number of optimal orderings grows exponentially, their proportion tends quickly to zero as $n$ grows. It is therefore, perhaps, surprising that we are able to find an optimal ordering so quickly from merely tens of shuffles. The proportion of optimal orderings (approximately the ratio of ordered trees to total trees, $\frac{(n-1)!}{(2n-3)!!}$, ranges from $8\times 10^{-4}$ in our smallest dataset (14 taxa) to $6\times 10^{-29}$ in the largest (99 taxa; see Table~\ref{tab:data}).

This efficiency comes from the topology-dependence of the Queue Shuffle algorithm, which allows us to plot a relatively ``smooth'' path through the space of possible orderings. That is, the majority of trees in the new ordered space will have similar properties to the previous optimal tree and so, unless the previous tree was a local minimum of the objective, it is likely that one of these ``close'' trees will have a lower objective value.

Lemma~\ref{lem:asym} shows that the expected distance from the root grows harmonically as the label increases. For large trees, the node with label $n-1$ has an expected distance from the root of approximately twice the expected distance from the root of the node label $0$. This property is noticeable even for small trees – if $n = 10$, then the ratio of the expected distance to the root of node 9 and node 0 is approximately $1.65$. Thus, nodes which are close to the root in the current optimum, will also be closer on average to the root in the new space of ordered trees. In essence, this means that “fewer slots are wasted'' in the new ordered space – that is, there are fewer trees in the new space of ordered trees which are topologically far from the previous optimum (a tree that, after the first few iterations, is likely to be far closer to the true optimum than a randomly-chosen tree) and hence, more trees which are reasonable candidates for having lower objective values.

Lemma~\ref{fig:swap} proves another example of the smoothness in transitions induced by the Queue Shuffle, based on nearest neighbour interchange (NNI) moves. An NNI move considers the four subtrees attached to two non-root nodes that share an edge and swaps two of these subtrees. Lemma~\ref{fig:swap} shows that, starting from a tree $\mathcal{T}$, any tree which is one NNI move away from $\mathcal{T}$ will be in the new space of ordered trees with probability at least $\frac{1}{4}$. 

This ensures that this new space  contains many sensible proposal trees. Perhaps the most surprising aspect of this theorem is that this probability is bounded below, independently of the topology. Thus, with high probability, the optimal tree will only remain the same for more than a few iterations if large sets of similar trees yield lower objective values than the current optimum. 

That being said, the Queue Shuffle does not guarantee that the global minimum will be found, even if the gradient-based algorithm for optimising $F(W)$ always converges to the optimal tree. If a tree is “far'' from the nearest tree with a better objective value, then it may take a very large number of shuffles (or, indeed, it may be impossible) to find a better tree. However, while the only theoretical guarantee is that Lemma~\ref{fig:swap} shows it will quickly find better trees that can be formed by NNI, we expect that stronger conditions hold on its ability to “escape'' from local minima.

\begin{table*}[htbp]
\centering
\caption{Evaluation datasets. rRNA/rDNA: ribosomal RNA/DNA, mtDNA: mitochondrial DNA. AA: amino acid. For the Jawed dataset, several subsets of the original dataset~\parencite{irisarri2017} were used (from 1,460 to 18,406 sites; cf. Fig.~\ref{fig:jawed_verts}).}
\begin{tabular}{lrrll}
\toprule
\textsb{Dataset} & \textsb{\# Sites} & \textsb{\# Taxa} & \textsb{Type} & \textsb{Taxonomic rank} \\
\small{(Reference)} \\
\midrule
DS1  & 1,949 & 27 & rRNA (18S) & Tetrapods \\ 
\small{\parencite{hedges1990}}\\
DS2  & 2,520 & 29 & rRNA (18S) & Acanthocephalans \\
\small{\parencite{garey1996}}\\
DS3  & 1,812 & 36 & mtDNA & Mammals; mainly Lemurs\\
\small{\parencite{yang2003}}\\
DS4  & 1,137 & 41 & rDNA (18S) & Fungi; mainly Ascomycota\\ 
\small{\parencite{henk2003}}\\
DS5  & 378 & 50 & DNA & Lepidoptera \\
\small{\parencite{brower2000, lakner2008}}\\
DS6  & 1,133 & 50 & rDNA (28S) & Fungi; mainly Diaporthales \\
\small{\parencite{zhang2001}}\\
DS7  & 1,824 & 59 & mtDNA & Mammals; mainly Lemurs\\
\small{\parencite{yoder2004}}\\
DS8  & 1,008 & 64 & rDNA (28S) & Fungi; mainly Hypocreales\\
\small{\parencite{rossman2001}}\\
DS9 & 955  & 67 & DNA & Poaecae (grasses)\\
\small{\parencite{ingram2004}}\\
DS10 & 1,098 & 67 & DNA & Fungi; mainly Ascomycota\\
\small{\parencite{suh1999}}\\
DS11 & 1,082 & 71 & DNA & Lichen \\
\small{\parencite{kroken2000}}\\
\midrule
Eutherian & 1,338,678 & 37 & DNA & Eutherian Mammals \\
\small{\parencite{Song2012-ql}}\\
Jawed & 1,460-18,406 & 99 & AA & Gnathostomata (jawed vertebrates) \\
\small{\parencite{irisarri2017}}\\
Primates & 232 & 14 & mtDNA & Mammals; mainly Primates \\
\small{\parencite{hasegawa1989}}\\
\small{\parencite{paradis2004}}\\
\bottomrule
\end{tabular}
\label{tab:data}
\end{table*}

\subsubsection*{Computational complexity}

The computational complexity for all distance-based algorithms requires an upfront computational cost of $\mathcal{O}(n^2)$ to compute the distance matrix. We will disregard this cost from subsequent comparisons. The standard neighbour-joining algorithm~\parencite{saitou1987} has an overall computational complexity of $\mathcal{O}(n^3)$. FastME has a computational complexity of $\mathcal{O}(k n^2 _ \text{Diam}(T))$ (where Diam$(T)$ is the maximum path length in a tree, which is generally much smaller than $n$) for $k$ iterations where $k<n$ when $n$ is large. When fully discrete, our algorithm also has same complexity but with added mechanisms for escaping optima via Queue Shuffle. Therefore, a discrete setting is as computationally efficient as FastME (see Appendix~\ref{append:est} for details). 

Computing the expectation in Algorithm~\ref{algo:get_edges_exp} has complexity $\mathcal{O}(n^3)$. A single gradient evaluation (that is, calculating $\frac{\partial F}{\partial W_{ij}}$ for some $i$ and $j$) is also $\mathcal{O}(n^3)$ and therefore computing the full Jacobian is $\mathcal{O}(n^5)$. Our Queue Shuffle algorithm runs in $\mathcal{O}(n)$. Therefore, our optimisation for $k$ steps and $l$ shuffles yields a complexity of $\mathcal{O}(kln^5)$. The size of $l$ is dependent on the choice of gradient optimiser, and the size of $k$ varies if a sensible ordering is initialised. 

Thus, the computational complexity of GradME is substantially higher than that of FastME and closer to that of FITCH~\parencite{felsenstein1997alternating}. This is due to the far greater mathematical complexity of the continuous objective function, $F(W)$. As it is an expectation over all possible ordered trees, the explicit formula for $F(W)$ is a polynomial in $W$ with $(n-1)!$ different terms. Intuitively, the continuous space always considers a path between any two trees, something that becomes impossible with discrete settings. Thus, being able to compute it in polynomial time is a vast improvement on a naive approach, although it is still considerably less than the innovative FastME greedy approach. More savings should be possible, we hope to make further efficiency gains in future work. 

\subsubsection*{Evaluation} We evaluate GradME on a diverse corpus of 14 empirical molecular sequence datasets (Table~\ref{tab:data}). The first 11 are commonly used to assess phylogenetic inference performance~\parencite{Whidden2015-jn}, whereas the last three were used to assess inference on rooted trees. For each dataset, we start from a random tree and optimise the $W$ matrix to a tolerance of 1e-10 using gradient descent with Adafactor~\parencite{shazeer2018} optimisation. The distance matrix $D$ is computed using the GTR+$\Gamma$ substitution model for DNA and an LG model~\parencite{Le2008-ut} for amino acids. Substitution model parameters for the GTR+$\Gamma$ are also estimated using gradient descent with Adafactor using a pairwise maximum likelihood approach~\parencite{yang2006}. Jukes-Cantor~\parencite{jukes1969}, F81~\parencite{felsenstein1981} and TN93~\parencite{tamura1993} models were also tested for DNA, while stochastic gradient descent (SGD), RMSprop~\parencite{tieleman2012}, and AdamW~\parencite{Kingma2014-ow, loshchilov2018} were also considered for optimisation (see Fig.~\ref{fig:convergence_analysis}). To fairly assess the performance of GradME, we compare our framework to two well-established distance-based methods: \texttt{BioNJ}~\parencite{gascuel1997}, based on the neighbour joining algorithm~\parencite{saitou1987}, and \texttt{FastME}~\parencite{lefort2015}, based on balanced minimum evolution.

\vspace{-1em}

\subsubsection*{Implementation} Implementation of the BME criterion and the optimisation framework was written in Python using \texttt{Jax}~\parencite{jax2018github} and \texttt{Optax}~\parencite{deepmind2020}. Optimisation was performed on a Xeon 2.30GHz (CPU; Intel Corporation) or on a single GeForce GTX 1080 (GPU; Nvidia Corporation). Evaluation of the \texttt{BioNJ}~\parencite{gascuel1997} and \texttt{FastME}~\parencite{lefort2015} methods was performed via the R package \texttt{ape}~\parencite{paradis2004} using \texttt{rpy2}~\parencite{rpy2}. Tree manipulation and visualisation scripts were written using \texttt{ete3}~\parencite{huerta2016} and \texttt{NetworkX}~\parencite{hagberg2008}. An implementation is available at: \url{https://github.com/Neclow/GradME}.

\vspace{-1em}

\section*{Data availability} \label{s:code_availability}
All code relevant to reproduce the experiments is available online: \url{https://github.com/Neclow/GradME}.


\begin{refcontext}
    \patchbibmacro{date+extradate}{%
      \printtext[parens]%
    }{%
      \setunit{\addperiod\space}%
      \printtext%
    }{}{}
    \printbibliography
\end{refcontext}

\section*{Author contributions}
S.B and M.J.P conceived of the study. S.B and C.A.D supervised. S.B, N.S, M.J.P and D.A.D designed the study. S.B and N.S performed optimisation runs. S.B, M.J.P, N.S and D.A.D performed analysis. All authors contributed to writing the original draft. M.J.P and J.P drafted the appendix. 

\section*{Competing interests}
The authors declare no competing interests.

\subsection*{Funding}
S.B. and C.A.D acknowledges support from the MRC Centre for Global Infectious Disease Analysis (MR/R015600/1), jointly funded by the UK Medical Research Council (MRC) and the UK Foreign, Commonwealth \& Development Office (FCDO), under the MRC/FCDO Concordat agreement, and also part of the EDCTP2 programme supported by the European Union.  SB is funded by the National Institute for Health Research (NIHR) Health Protection Research Unit (HPRU) in Modelling and Health Economics, a partnership between the UK Health Security Agency, Imperial College London and LSHTM (grant code NIHR200908). Disclaimer: “The views expressed are those of the author(s) and not necessarily those of the NIHR, UK Health Security Agency or the Department of Health and Social Care.” S.B. acknowledges support from the Novo Nordisk Foundation via The Novo Nordisk Young Investigator Award (NNF20OC0059309).  S.B. acknowledges support from the Danish National Research Foundation via a chair grant which also supports N.S. S.B. acknowledges support from The Eric and Wendy Schmidt Fund For Strategic Innovation via the Schmidt Polymath Award (G-22-63345). S.B and N.S acknowledge the Pioneer Centre for AI, DNRF grant number P1 as affiliate researchers. C.A.D receives support from the NIHR HPRU in Emerging and Zoonotic Infections, a partnership between the UK Health Security Agency, University of Liverpool, University of Oxford and Liverpool School of Tropical Medicine (grant code NIHR200907). D.A.D. is funded by a European Research Council Marie Sklodowska-Curie fellowship (H2020-MSCA-IF-2019-883832). M.P. was funded by a DTP Studentship from the Engineering and Physical Sciences Research (EPSRC)

\setcounter{figure}{0}
\setcounter{table}{0}
\makeatletter
\renewcommand{\thefigure}{S\@arabic\c@figure}
\renewcommand{\thetable}{S\@arabic\c@table}
\makeatother


\section*{Appendix}

\subsection{BME for rooted trees}\label{sec:bme_rooted}
\subsubsection*{Comparing the rooted and unrooted objectives}
\begin{lemma}
\label{lem:rooted1}
    Consider adding an extra taxon, $n$, to the set of taxa such that, for some $D^*$ and $\delta$.
    \begin{equation}
    |D_{ni} - D^*| < \delta \quad \forall i \neq n
\end{equation}
This creates an unrooted tree $\mathcal{T}^u$, and we can create a rooted tree $\mathcal{T}^r$ by removing node $n$. If $e^u_{ij}$ denotes inter-taxa distance in $\mathcal{T}^u$ and $e_{ij}^r$ denotes inter-taxa distance in $\mathcal{T}^r$, then
\begin{equation}
    \bigg|\sum_{i=0}^n\sum_{j=0}^nD_{ij}2^{-e^u_{ij}} - \sum_{i=0}^{n-1}\sum_{j=0}^{n-1}D_{ij}2^{-e^r_{ij}} - D^*\bigg| \leq \delta
\end{equation}
\end{lemma}
\textbf{Proof:} Firstly, note that for $i,j < n$,
\begin{equation}
    e_{ij}^r = e_{ij}^u
\end{equation}
as the path between $i$ and $j$ will not contain any leaf nodes other than $i$ and $j$, and therefore will not contain node $n$. Then, using the fact that $D_{nn} = 0$,
\begin{align}
    \sum_{i=0}^n\sum_{j=0}^nD_{ij}2^{-e^u_{ij}} &= \sum_{j=0}^{n-1}D_{nj}2^{-e^u_{nj}} + \sum_{i=0}^{n-1}D_{in}2^{-e^u_{in}} + \sum_{i=0}^{n-1}\sum_{j=0}^{n-1}D_{ij}2^{-e^u_{ij}}\\
    &\leq \sum_{j=0}^{n-1}(D^* + \delta)2^{-e_{nj}} +\sum_{i=0}^{n-1}(D^*+ \delta)2^{-e^u_{in}} + \sum_{i=0}^{n-1}\sum_{j=0}^{n-1}D_{ij}2^{-e^r_{ij}}
\end{align}
We can make progress with this sum by noting the Kraft Equality, found throughout the literature~\parencite{catanzaro2012} 
\begin{equation}
    \sum_{j}2^{-e^u_{nj}} = \frac{1}{2}
\end{equation}
This means that
\begin{equation}
     \sum_{i=0}^n\sum_{j=0}^nD_{ij}2^{-e^u_{ij}}  \leq \sum_{i=0}^{n-1}\sum_{j=0}^{n-1}D_{ij}2^{-e^r_{ij}} + D^* + \delta
\end{equation}
Similarly, one can show
\begin{equation}
    \sum_{i=0}^n\sum_{j=0}^nD_{ij}2^{-e^u_{ij}}  \geq \sum_{i=0}^{n-1}\sum_{j=0}^{n-1}D_{ij}2^{-e^r_{ij}} + D^* - \delta
\end{equation}
and hence the result follows.
\subsubsection*{Understanding the BME rooting}
\begin{definition}
\label{def:distance}
    \textbf{Distance to root heuristic:}  Under the assumption that the tree is ultrametric, we estimate the distance between the root and the leaf taxa using the following algorithm:

    \textbf{1)} Find two leaf nodes that share a parent that is not the root. If no such pair exists, then move to step 3.

    \textbf{2)}  Let $d_1$ and $d_2$ be the distance between each leaf node and its parent. Suppose that $d_3$ is the distance between the parent and its parent (i.e. the grandparent of the leaves). Remove the leaves from the tree, update the distance between the parent and its parent to be $d_3 + \frac{d_1 + d_2}{2}$ and return to step 1

    \textbf{3)} If $d_1$ and $d_2$ are the distances between the two remaining children and the root, then the distance between the root and the taxa is $\frac{d_1+d_2}{2}$.
\end{definition}

\begin{lemma}
\label{lem:root_min}
    Consider the true unrooted tree $\mathcal{T}$ and suppose that the distance matrix $D$ gives the true (time) distances between each pair of taxa in $\mathcal{T}$. 

    Then, the optimal rooting — that is, the edge in the tree such that placing the root on this edge minimizes the BME objective — minimizes the distance to root heuristic defined in Definition \ref{def:distance}.
\end{lemma}
\textbf{Proof:} Choose an edge, $b$ on which to place the root. We define an indicator function $X_b$ such that $X_b(A,B)$ is 1 if $b$ is on the path between nodes $A$ and $B$ and 0 otherwise. Adding a root to edge $b$ changes the objective function by halving the weight assigned to each $D_{AB}$ such that $X_b(A,B) = 1$ as the path length between these nodes will increase by 1. Thus, using $f_r$ to denote the rooted objective and $f_u$ to denote the unrooted objective,
\begin{equation}
    f_r = f_u - \frac{1}{2}\sum_{X_b(A,B) = 1}D_{AB}2^{-e_{AB}}
\end{equation}
where the $e_{AB}$ are the path lengths in the original, unrooted tree and both $A$ and $B$ are allowed to vary in the sum. Hence, for a fixed unrooted tree topology, the optimal rooting will solve
\begin{equation}
\label{eq:rooting}
    \max_b\bigg\{\sum_{X_b(A,B) = 1}D_{AB}2^{-e_{AB}}\bigg\}
\end{equation}
By Lemma \ref{lem:supp_1},
\begin{equation}
\label{eq:root_reform2}
    \sum_{X_b(A,B) = 1}D_{AB}2^{-e_{AB}} = 2\sum_A2^{-g_A}D_{Ar}
\end{equation}
while, by Lemma \ref{lem:supp_2}, the root-to-tip heuristic, $\mathcal{D}$, satisfies
\begin{equation}
    \mathcal{D} = \sum_A2^{-g_A}D_{Ar}
\end{equation}
and hence, the optimal rooting solves
\begin{equation}
    \max_b\bigg\{\mathcal{D}\bigg\}
\end{equation}
as required.
\subsection{Ordered Trees}\label{sec:ordered}

\begin{definition}
\label{def:ltr}
    \textbf{Left-to-right construction algorithm} An ordered tree can be constructed as follows:

    \textbf{1)} Begin with nodes $0$ and $1$, each joined to a root. Label the edge joining node $0$ to the root as edge $0$, and the edge joining node $1$ to the root as edge $1$

    \textbf{2)} Process the nodes in order 2, 3, 4… When processing node $k$, join it to edge $v_k$ (creating a new internal node — this is well-defined as $v_k < k$). Label the edge joining node $k$ to the tree as edge $k$.
    
\end{definition}
\begin{lemma}
\label{lem:ltr}
    The left-to-right algorithm in Definition \ref{def:ltr} and the standard Phylo2Vec algorithm defined in \parencite{penn2023} produce the same tree, provided $\boldsymbol{v}$ is ordered.
\end{lemma}
\textbf{Proof:} To show that the left-to-right algorithm gives an equivalent tree, define $\mathcal{T}$ to be the tree resulting from the standard Phylo2Vec algorithm and $\mathcal{T}'$ to be the tree resulting from this new left-to-right algorithm. We proceed by induction on the number of nodes, $n$, noting that the case $n=2$ is trivial.

Suppose now that the algorithms are equivalent for $n = m$. Choose some ordered $\boldsymbol{v}$ of length $m+1$ (so that this corresponds to $n = m+1$) and consider processing the first node using the Phylo2Vec algorithm, so that (using the fact that $\boldsymbol{v}$ is ordered so that no nodes are skipped), node $m$ merges with node $v_m$. 

From this step, the Phylo2Vec algorithm proceeds as if there were $n-1$ nodes and the vector was $\tilde{\boldsymbol{v}} = (v_0,v_1,...,v_{m-1})$. Define $\tilde{\mathcal{T}}$ to be the tree given by $\tilde{\boldsymbol{v}}$. Hence, $\tilde{\mathcal{T}}$ can be created from $\mathcal{T}$ by removing nodes $v_{m}$ and $m$ (and the edges connecting them to the tree) and relabelling their parent as $v_m$. Equivalently (by reversing this process), $\mathcal{T}$ can be created from $\tilde{\mathcal{T}}$ by adding a node to the edge joining leaf node $v_m$ to the tree, and connecting node $m$ to this edge.

Moreover, from the inductive hypothesis, $\tilde{\mathcal{T}}$ can be constructed by using the left-to-right algorithm on $\tilde{\boldsymbol{v}}$. As this algorithm processes $v_m$ last, this means that after $m$ nodes have been added to the tree by the left-to-right algorithm applied to $\boldsymbol{v}$, the current tree is given by $\tilde{\mathcal{T}}$. 
\\
\\
\noindent
The final step of the left-to-right algorithm applied to $\boldsymbol{v}$ is to add a node to the edge joining leaf node $v_m$ to the tree, and to connect node $m$ to this edge. As previously discussed, this creates the tree $\mathcal{T}$ and hence, $\mathcal{T} = \mathcal{T}'$ as required.

\subsection{A continuous objective function}\label{sec:cont_edge}
\subsubsection*{Construction}
\begin{lemma}
\label{lem:cont_obj}
    For a randomly chosen tree with distribution $W$, define, for $i,j,<k$, $e_{ij}^k$ to be the path length between taxa $i$ and $j$ when $k$ nodes have been added to the tree (using the left-to-right construction algorithm). Define $E_{ik}^k := \mathbb{E}(2^{-e_{ij}^k})$. Then, for $i < j$
    \begin{equation}
        E_{ij}^{k+1} = \left\{\begin{matrix}
            E^{k}_{ij} \bigg[1 - \frac{1}{2}(W_{ki} + W_{kj})\bigg] & \text{if $i < j < k$}\\
       \bigg[\frac{1}{2}\sum_{x\neq i}E_{ix}^kW_{kx}\bigg] + \frac{1}{4}W_{ki} & \text{if $i < k$}
        \end{matrix}\right.
    \end{equation}
    with the remaining values following by symmetry.
\end{lemma}
\textbf{Proof:} Adding node $k$ to the tree increases the path length between nodes $i$ and $j$ by 1 if and only if $V_k = i$ or $V_k = j$. As this condition is independent of other values of $\boldsymbol{V}$, using $e_{ij}^k$ to be the path length after the $k$ nodes  $\{0,...,k-1\}$ have been added, one can write
\begin{equation}
    2^{-e^{k+1}_{ij}} =  2^{-(e^{k}_{ij} + \mathds{I}\{V_k \in \{i,j\}\})} =     2^{-e^{k}_{ij}}\times 2^{-\mathds{I}\{V_k \in \{i,j\}\}}
\end{equation}
This is a product of independent random variables and so, defining $E^k_{ij} := \mathbb{E}(2^{-e^k_{ij}})$ and noting that $\mathds{I}\{V_k \in \{i,j\}\}$ is a Bernoulli random variable with probability $W_{ki} + W_{kj}$, this equation becomes
\begin{equation}
\label{eq:iter}
    E^{k+1}_{ij} = E^{k}_{ij} \bigg[(1 - (W_{ki} + W_{kj}) + \frac{1}{2}(W_{ki} + W_{kj})\bigg] =  E^{k}_{ij} \bigg[1 - \frac{1}{2}(W_{ki} + W_{kj})\bigg]
\end{equation}
To close this iterative system, note that when node $j$ is added to the tree, it will be a path length 2 from the leaf node $V_j$ connecting the edge it is joined to. Moreover, the distance between node $j$ and any other nodes $x\neq V_j$ in the tree will be equal to one plus the distance between node $x$ and node $V_j$. That is,
\begin{equation}
    2^{-e_{ij}^{j+1}} = \left\{\begin{matrix} 2^{-(1+e_{i,V_j}^j)} & \text{if $V_j \neq i$} \\
    \frac{1}{4} & \text{if $V_j = i$} \end{matrix} \right.
\end{equation}
Thus, conditioning on the value of $V_j$,
\begin{equation}
\label{eq:init}
    E_{ij}^{j+1} = \bigg[\frac{1}{2}\sum_{x\neq i}E_{ix}^jW_{jx}\bigg] + \frac{1}{4}W_{ji}
\end{equation}
Finally, by symmetry, $E_{ji}^{j+1} = E_{ij}^{j+1}$. Noting that $E_{ij}^k$ is undefined (and unnecessary) for $k < \max(i,j) + 1$, as nodes $i$ and $j$ have not both been added to the tree, (\ref{eq:iter}) and (\ref{eq:init}) hence form a closed system. This can be solved inductively, finding all $E_{ij}^m$ terms for $m=2,3,...,n$.
\subsubsection*{Discrete minima}
\begin{lemma}
\label{lem:discrete}
    Define $f(\boldsymbol{v})$ to be the objective function for the discrete tree given by the Phylo2Vec vector $\boldsymbol{v}$. Then, if $V^*$ is the set of vectors $\boldsymbol{v}$ which minimize f, any optimal $W$ satisfies
    \begin{equation}
        \mathbb{P}(\boldsymbol{V} \in V^*|W) = 1
    \end{equation}
     Moreover if $|V^*| = 1$ then there is a unique $\boldsymbol{v}$ minimizing $f$, then there is a unique $W$ minimizing $F$, which is the matrix such that
\end{lemma}
\textbf{Proof:} Define $V$ to be the set of ordered tree vectors. Then,
\begin{align}
       \label{eq:probeq}  F(W) = \sum_{\boldsymbol{u} \in V}\mathbb{P}(\boldsymbol{V} = \boldsymbol{u}|W)f(\boldsymbol{u})
= \sum_{\boldsymbol{u} \in V}\prod_{m=1}^{n-1}W_{m,u_m}f(\boldsymbol{u})
\end{align}
As
\begin{equation}
     \sum_{\boldsymbol{u} \in V}\mathbb{P}(\boldsymbol{V} = \boldsymbol{u}|W) = 1
\end{equation}
we see that $F(W)$ is a weighted average of the values of $f(\boldsymbol{u})$ for $\boldsymbol{u} \in V$. Thus, for any $\boldsymbol{v} \in V^*$
\begin{equation}
     F(W)  \geq f(\boldsymbol{v})
\end{equation}
and
\begin{equation}
    F(W) =  f(\boldsymbol{v}) \quad \Rightarrow \quad \mathbb{P}(\boldsymbol{V} \in V^*|W) = 1
\end{equation}
as required. If $V = \{\boldsymbol{v}\}$, this minimum requires
\begin{equation}
     \mathbb{P}(\boldsymbol{V} = \boldsymbol{v}|W) = 1
\end{equation}
 and therefore, using (\ref{eq:probeq}) $W$ is uniquely defined by
 \begin{equation}
     W_{m,j} = \mathds{I}\{v_m = j\} \quad \forall m,j
 \end{equation}
 as required.
\subsection{Orderings}
This section considers the labelling algorithm introduced in the main text.
\begin{lemma}
\label{lem:ordering_1}
    Two nodes have the same label only if they share an ancestor with that label.
\end{lemma}

\textsb{Proof:} Suppose that this is false, and that nodes $x$ and $y$ have the same label, $L$, but do not share an ancestor with that label. Define $a(x)$ and $a(y)$ to be the nodes of lowest generation (that is, the nodes closest to the root) with label $L$ such that they are ancestors of $x$ and $y$ respectively. Note that, by assumption, $a(x) \neq a(y)$ and also, neither can be the root (as the root is the ancestor of all nodes). Moreover, they cannot share a parent, as the children of a parent are labelled differently. Thus, without loss of generality, one can assume that $a(x)$ was labelled first. By definition, the parent of $a(x)$ does not have label $L$ and hence, $L$ must have been the smallest unused label when node $a(x)$ was labelled. A similar argument for $a(y)$ shows that $L$ must have been the smallest unused label when node $p(y)$ was labelled. However, when node $a(y)$ was labelled, $L$ had been used to label $a(x)$, giving the required contradiction.
\begin{lemma}
\label{lem:ordering_2}
    $l$ is a permutation of the set $\{0,1,...,n-1\}$.
\end{lemma} 
\textbf{Proof:} Firstly, note that as there are $n-1$ internal nodes, and a single new label is introduced every time the children of an internal node are labelled, the set of labels used (across all nodes) must be $\{0,1,...,n-1\}$.
\\
\\
\noindent
Suppose that leaf nodes $x$ and $y$ have the same label $L$. By Lemma~\ref{lem:ordering_1}, they must share an ancestor $a$ with that label. Define $b$ to be the shared ancestor with the highest generation (that is, furthest from the root) such that $b$ has label $L$. Then, either nodes $x$ and $y$ are the children of $b$ (in which case, they have distinct labels) or they are descendants of distinct children, $c$ and $d$, of $b$. In the second case, one can impose without loss of generality that $c$ does not have label $L$ and, as label $L$ has already been used to label $b$, we know that all descendants of $c$ do not have label $L$. Hence, in both cases, one of $x$ and $y$ does not have label $L$ as required.
\begin{lemma}
\label{lem:ordering_3}
    For each $i \in \{0,1,...,n-1\}$, define $x(i)$ to be the node of the highest generation with label $i$. Define (for $i > 0$), $y(i)$ to be the label of the parent of $x(i)$. Then, with ordering $l$, the tree is given by
\begin{equation}
    v_0 = 0 \quad \text{and} \quad v_{i} = y(i) \quad \forall i > 0
\end{equation}
\end{lemma} 
\textbf{Proof:} Firstly, note that $\boldsymbol{v}$ is ordered, as the label of a child is greater than or equal to that of its parent. Hence, as the parent of $x(i)$ does not have label $i$, it must have a label strictly less than $i$ and so $v_i < i$ as required.
\\
\\
\noindent
Consider constructing the tree according the “standard'' right-to-left Phylo2Vec algorithm. One can then proceed by induction on the number of nodes. The case $n=2$ is trivial, and so suppose it holds for $n = m$ and consider a tree with $n = m+1$.
\\
\\
\noindent
The first node to be processed has label $m$. No other node can have label $m$ cannot also have label $m$ (as, otherwise, one of its children would have label greater than $m$, contradicting Lemma \ref{lem:ordering_2}) and hence, the first node to be processed is $x(m)$. The parent of $x(m)$ must have label $y(m)$.  This must also be the label of its other child and hence, $v_m = y(m)$ ensures that the node with label $m$ merges with its sibling from the original tree (which is correct). 
\\
\\
\noindent
From this point, the tree now has $m$ nodes, and our right-to-left construction algorithm considers the parent of the node labelled $m$ to now be a leaf node with label $y(m)$. The values of $y(i)$ for this tree are unchanged (in particular, $y(y(m))$ depends on the label of an ancestor of the parent of node $m$) and hence, by induction, the remaining values of $\boldsymbol{v}$ correctly generate the rest of the tree. Thus, the correct tree is generated by $\boldsymbol{v}$ as required.
\subsection{Queue Shuffle: generating principled ordering proposals}\label{sec:queue_shuffle}
\subsubsection*{Asymmetry of ordered tree spaces}
\begin{lemma}
\label{lem:asym}
 Define $g_m^k$ to be the expected distance from the root of the node labelled $m$ in a tree with $k$ nodes. Define the harmonic sum function
 \begin{equation}
      H(m) = \left\{ \begin{matrix} \sum_{j=1}^m\frac{1}{j} &\text{if $m \geq 1$}\\ 1 & \text{if $m=0$}\end{matrix}\right.
 \end{equation}
 Then,
    \begin{equation}
    \label{eq:claimed_sol}
    g^k_m =H(k-1) + H(m) - 1
\end{equation}
\end{lemma}
\textbf{Proof:} From our left-to-right construction algorithm,
\begin{equation}
\label{eq:init_eq}
    g^k_{k-1} = \sum_{m=0}^{k-2}\frac{1}{k-1} (1+g^{k-1}_m)
\end{equation}
as adding node $k-1$ according to $v_{k-1} = m$ means that the path length between node $k-1$ and the root will be one more than the path length of between $m$ and the root. If $v_{k-1} = m$, it will also increase the path length between node $m$ and the root by 1 and so
\begin{equation}
    g^k_m = \frac{k-2}{k-1}g^{k-1}_m + \frac{1}{k-1}(g^{k-1}_m + 1) = g^{k-1}_m + \frac{1}{k-1}
\end{equation}
The initial conditions of this system are that
\begin{equation}
    g_0^2 = g_1^2 = 1
\end{equation}
as in a two-node rooted tree, the leaves are distance 1 from the root. 

We claim by induction on $k$ that the solution to this system is (\ref{eq:claimed_sol}).

Note that this holds for $k = 2$ as $H(1) = H(0) = 1$. Moreover, under the inductive hypothesis that it holds for a tree with $k-1$ nodes, for $m < k-1$,
\begin{equation}
     g^k_m = g^{k-1}_m + \frac{1}{k-1} =  H(k-2) + H(m) - 1 + \frac{1}{k-1} = H(k-1) + H(m) - 1
\end{equation}
and
\begin{align}
    g^k_{k-1} &= \sum_{m=0}^{k-2}\frac{1}{k-1} (1+g^{k-1}_m)\\
    &= \sum_{m=0}^{k-2}\frac{1}{k-1} (1 + H(k-2) +H(m) - 1)\\
    &= \sum_{m=0}^{k-2}\frac{1}{k-1}(H(k-2) + H(m))\\
    &= H(k-2) + \frac{1}{k-1}\sum_{m=0}^{k-2}H(m)\\
    &= H(k-1) + \frac{1}{k-1}\sum_{m=1}^{k-2}\sum_{j=1}^m\frac{1}{j}
\end{align}
Now,
\begin{equation}
    \sum_{m=1}^{k-2}\sum_{j=1}^m\frac{1}{m} = \sum_{j=1}^{k-2}\sum_{m = j}^{k-2}\frac{1}{j} = \sum_{j=1}^{k-2}\frac{k-1-j}{j} = (k-1)H(k-2) -(k-2)
\end{equation}
and hence
\begin{equation}
     g^k_{k-1} = H(k-1) + H(k-2) -\frac{(k-2)}{(k-1)} = H(k-1) + H(k-1) - 1
\end{equation}
as required.
\subsubsection*{Nearest Neighbour Interchange}
\begin{definition}
    Define $\tau(\sigma)$ to be the space of possible trees given an ordering $\sigma$. 
\end{definition} 
\begin{definition}
For a tree $\mathcal{T}$, define $Q(\mathcal{T})$ to be the random ordering generated by Queue Shuffle.
\end{definition}
\begin{lemma}
\label{lem:swap}
    Consider a tree $\mathcal{T}$ and suppose that another tree, $\mathcal{T}'$ is one NNI move away from $\mathcal{T}'$. Then
\begin{equation}
    \mathbb{P}\bigg[\mathcal{T}'\in \tau\bigg(Q(\mathcal{T})\bigg)\bigg] \geq \frac{1}{4}
\end{equation}
\end{lemma}
 To facilitate the proof, we first note that for any permutation $\sigma$, by Lemma \ref{lem:supswap1}, that
\begin{equation}
    \mathbb{P}\bigg[Q(\mathcal{T}) = \sigma\bigg] \in \bigg\{0,\frac{1}{2^{n-1}}\bigg\}
\end{equation}
\begin{figure*}[htbp]
    \centering
    \includegraphics[width=0.8 \textwidth]{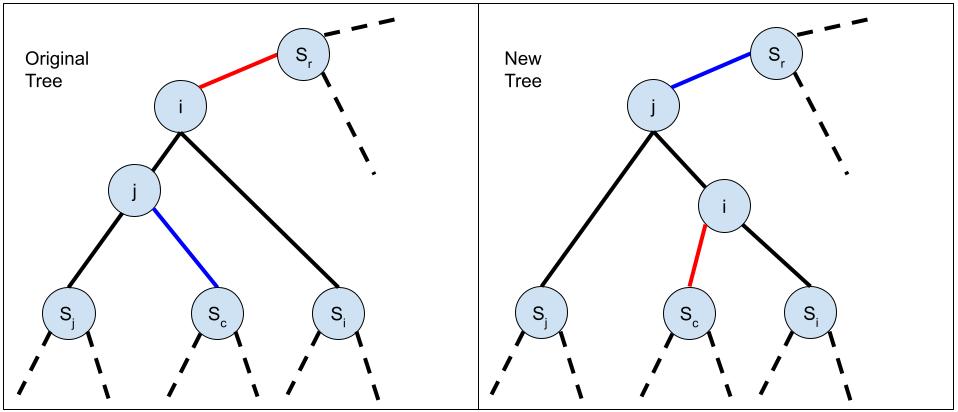} %
    \caption{A simple representation of the swapping property of Queue Shuffle. The swapped subtrees have roots labelled $S_i$ and $S_j$ while the node label $c$ is the subject of the equivalent subtree-prune and regraft operation. Note that in step 4, the internal nodes are renamed to illustrate that the required swap has indeed occurred.}
    \label{fig:swap}
\end{figure*}
\noindent
Now, we consider labelling the tree as shown in the left panel of Fig.~\ref{fig:swap}. Suppose that the edge to which the four subtrees are joined has nodes $i$ and $j$, with node $i$ being a higher generation than node $j$. Suppose that the subtrees rooted at the child of $j$ are labelled as $\mathcal{S}_j$ and $\mathcal{S}_c$. Suppose that the other child of $i$ (that is, the child not equal to $j$) is the root of a subtree $\mathcal{S}_i$. Finally, suppose that the fourth subtree is labelled $\mathcal{S}_r$ (this is the subtree containing the root). A single NNI move in this context involves swapping a pair of subtrees from the set $\{\mathcal{S}_i, \mathcal{S}_j,\mathcal{S}_r, \mathcal{S}_c\}$.

By symmetry, swapping the pair $\mathcal{S}_j$ and $\mathcal{S}_c$ or the pair $\mathcal{S}_r$ and $\mathcal{S}_i$ has no impact, as the new tree will be topologically equivalent to $\mathcal{T}$ (which is always in the new shuffled space). Thus, one needs only to prove the result for the tree formed by swapping $\mathcal{S}_c$ and $\mathcal{S}_r$. This is topologically equivalent to the tree in the right panel of Fig.~\ref{fig:swap}. We assume that this tree is $\mathcal{T}'$

 For a given permutation $\sigma$, define $l_i({\sigma})$ to be the label assigned to node $i$, $l_j({\sigma})$ to be the label assigned to node $j$,  $l^{\mathcal{S}}_i({\sigma})$ to be the label assigned to the root of $\mathcal{S}_i$, $l^{\mathcal{S}}_j({\sigma})$ to be the label assigned to the root of $\mathcal{S}_j$ and $l^{\mathcal{S}}_c(\sigma)$ to be the label assigned to the root of $\mathcal{S}_c$

Then, by Lemma \ref{lem:supswap2},
    \begin{equation}
    \mathbb{P}\bigg[l^{\mathcal{S}}_j({Q(\mathcal{T})})  =l_i({Q(\mathcal{T})})  \bigg] = \frac{1}{4}
\end{equation}

Now, suppose for a permutation $\sigma$ that $l^{\mathcal{S}}_j(\sigma)  =l_i(\sigma)$. Necessarily (as $r_B = 0$), $l^{\mathcal{S}}_c(\sigma)$ was the smallest available label when processing node $j$, which must be bigger than the smallest available label when node $i$ was processed (as $j$ is the child of $i$ and hence processed later). Thus, $l^{\mathcal{S}}_c(\sigma) >l^{\mathcal{S}}_i(\sigma)$. 
\\
\\
\noindent
A further important result proved in Lemma \ref{lem:supswap2} is that node $j$ must have been placed ahead of the root of $\mathcal{S}_i$ in the queue. Thus, all nodes in $\mathcal{S}_i$ are labelled either as $l^{\mathcal{S}}_i(\sigma)$ or with a label that is greater than $l^{\mathcal{S}}_c(\sigma)$, as all internal nodes in $\mathcal{S}_i$ are processed after node $j$. Moreover, all nodes in $\mathcal{S}_j$ are labelled as either $l^{\mathcal{S}}_j(\sigma)$ or a label greater than $l_c(\sigma)$ as they were processed after node $j$.
\\
\\
\noindent
Define  $\boldsymbol{v}$ to be the vector giving $\mathcal{T}$ under $\sigma$. Then, define a new vector $\boldsymbol{v}'$ by
\begin{equation}
    v'_{m} =\left\{ \begin{matrix} l^{\mathcal{S}}_i(\sigma) & \text{if $m =l^{\mathcal{S}}_c(\sigma) $}\\
    v_m & \text{otherwise} \\
    \end{matrix}\right.
\end{equation}
Then, $\boldsymbol{v}'$ is ordered as $l^{\mathcal{S}}_c(\sigma) > l_j^{\mathcal{S}}(\sigma)$. We now show that $\boldsymbol{v}'$ generates $\mathcal{T}'$ by using the left-to-right construction algorithm. Note that up to the point that the node labelled $l^{\mathcal{S}}_c(\sigma)$ is processed by this algorithm, no nodes have appended to the edges connecting either the root of $\mathcal{S}_i$ or $\mathcal{S}_j$ to the tree. This holds because all other nodes in $\mathcal{S}_i$ and $\mathcal{S}_j$ have labels greater than $l^{\mathcal{S}}_c(\sigma)$. Changing from $\boldsymbol{v}$ to $\boldsymbol{v}'$ means that the leaf node labelled $l^{\mathcal{S}}_c(\sigma)$ is joined to the edge connecting the root of $\mathcal{S}_i$ rather than the root of $\mathcal{S}_j$. The rest of the tree is then constructed identically as the vectors $\boldsymbol{v}$ and $\boldsymbol{v}'$ are the same.
\\
\\
\noindent
Thus, changing from $\boldsymbol{v}$ to $\boldsymbol{v}'$ generates the tree in the right panel of Fig.~\ref{fig:swap}, and therefore the tree formed by the NNI move swapping the trees $\mathcal{S}_c$ and $\mathcal{S}_r$.
\\
\\
\noindent
This completes the proof as it shows that
\begin{equation}
     \mathbb{P}\bigg[\mathcal{T}'\in \tau\bigg(Q(\mathcal{T})\bigg)\bigg| l^{\mathcal{S}}_j(\sigma)  =l_i(\sigma)\bigg] =1
\end{equation}
and hence, by Lemma~\ref{lem:supswap2}
\begin{equation}
    \mathbb{P}\bigg[\mathcal{T}'\in \tau\bigg(Q(\mathcal{T})\bigg)\bigg] \geq  \mathbb{P}\bigg[\mathcal{T}'\in \tau\bigg(Q(\mathcal{T})\bigg)\bigg| l^{\mathcal{S}}_j(\sigma)  =l_i(\sigma)\bigg]\mathbb{P}\bigg[l^{\mathcal{S}}_j(\sigma)  =l_i(\sigma)\bigg] = \frac{1}{4}
\end{equation}

\clearpage
\subsection{Eutherian Mammal phylogeny~\parencite{Song2012-ql}}
\FloatBarrier
\begin{figure}[!htb]
    \centering
    \includegraphics[scale=0.5]{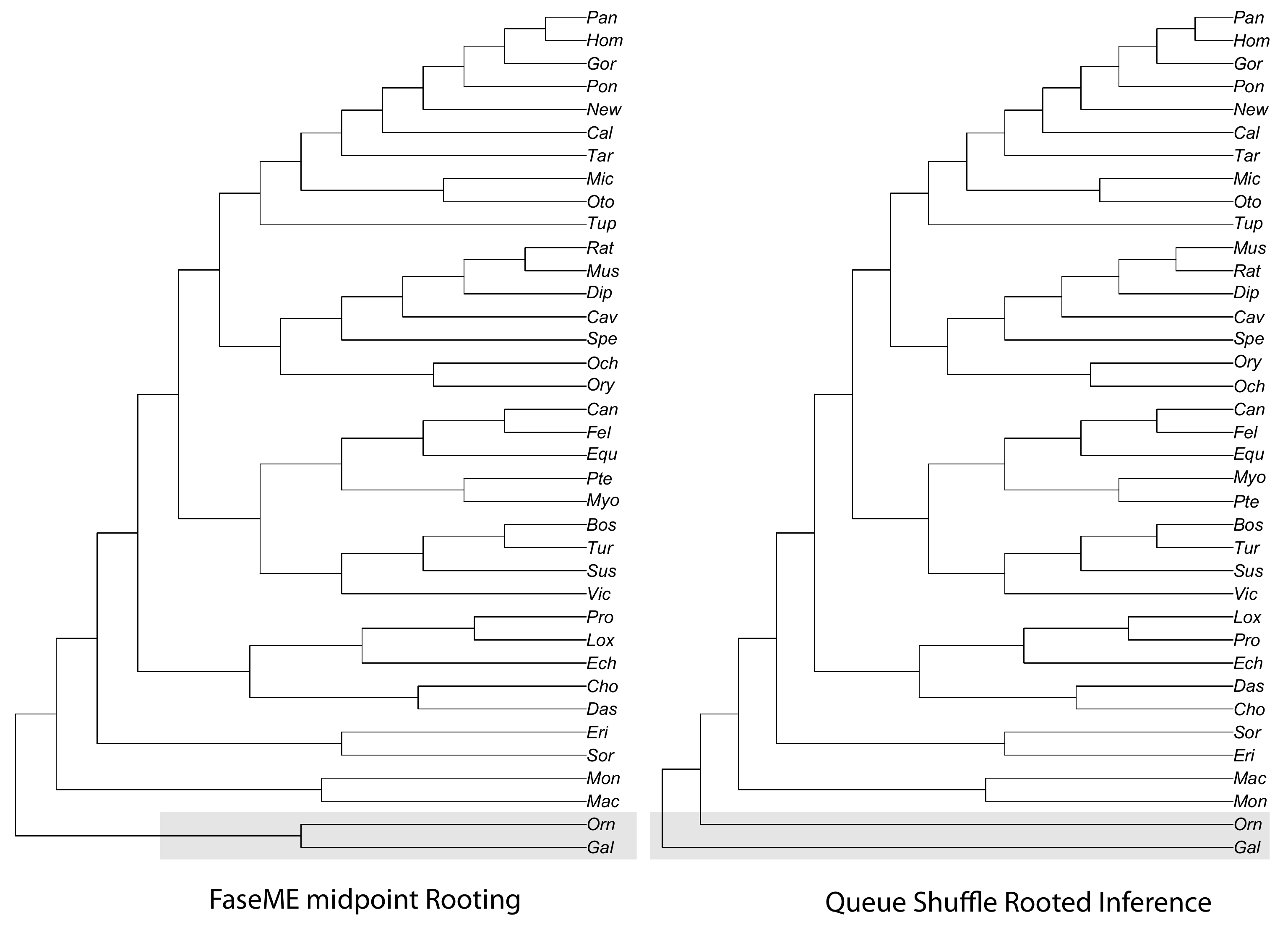}
    \caption{Comparison of the best unrooted FastME tree that has been midpoint rooted to an optimised rooted tree via Queue Shuffle on an Eutherian mammal dataset~\cite{Song2012-ql}. Queue Shuffle correctly places \emph{Gallus gallus} as the outgroup of mammals. Branch lengths are ignored and trees are displayed as ultrametric.}
    \label{fig:Songtree}
\end{figure}

\clearpage
\subsection{Convergence analysis}
\begin{figure*}[htbp]
    \centering
    \captionsetup[subfigure]{justification=raggedright,singlelinecheck=false}
    \begin{subfigure}{\textwidth}
        \caption{}
        \centering
        \includegraphics[width=0.8 \textwidth]{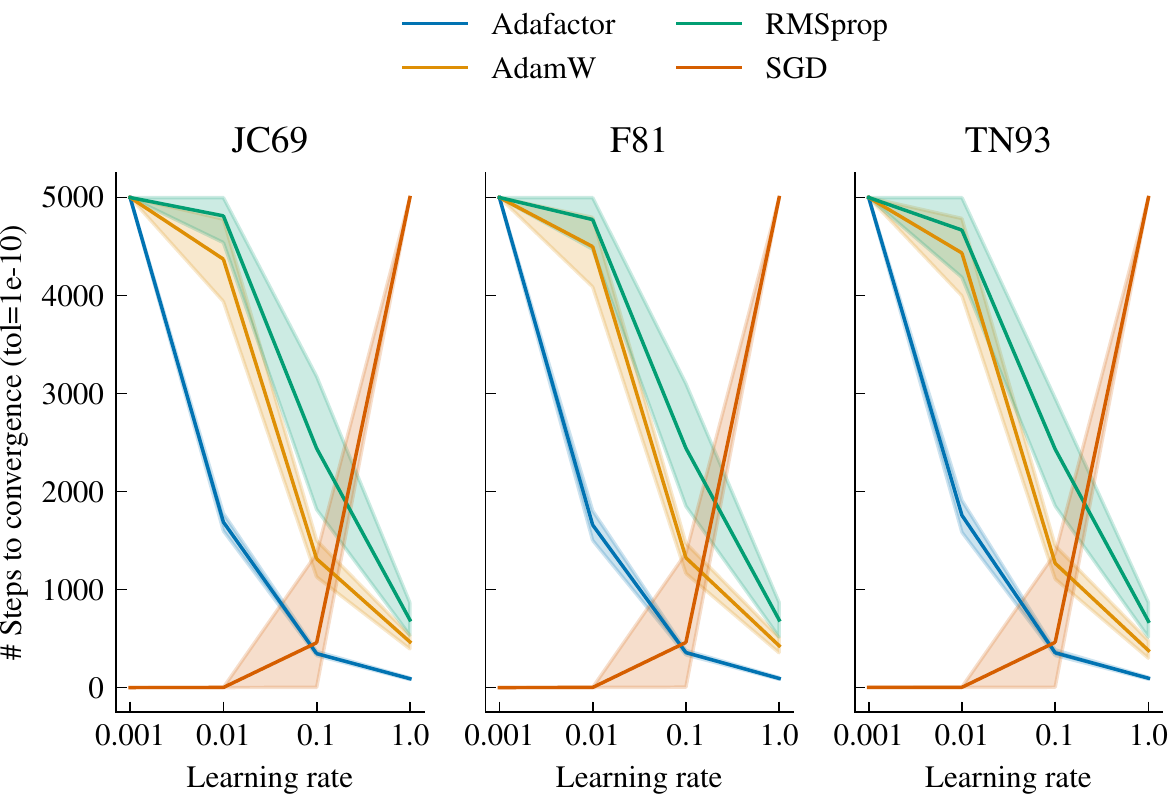}
    \end{subfigure}
    \begin{subfigure}{\textwidth}
        \caption{}
        \centering
        \includegraphics[width=0.8 \textwidth]{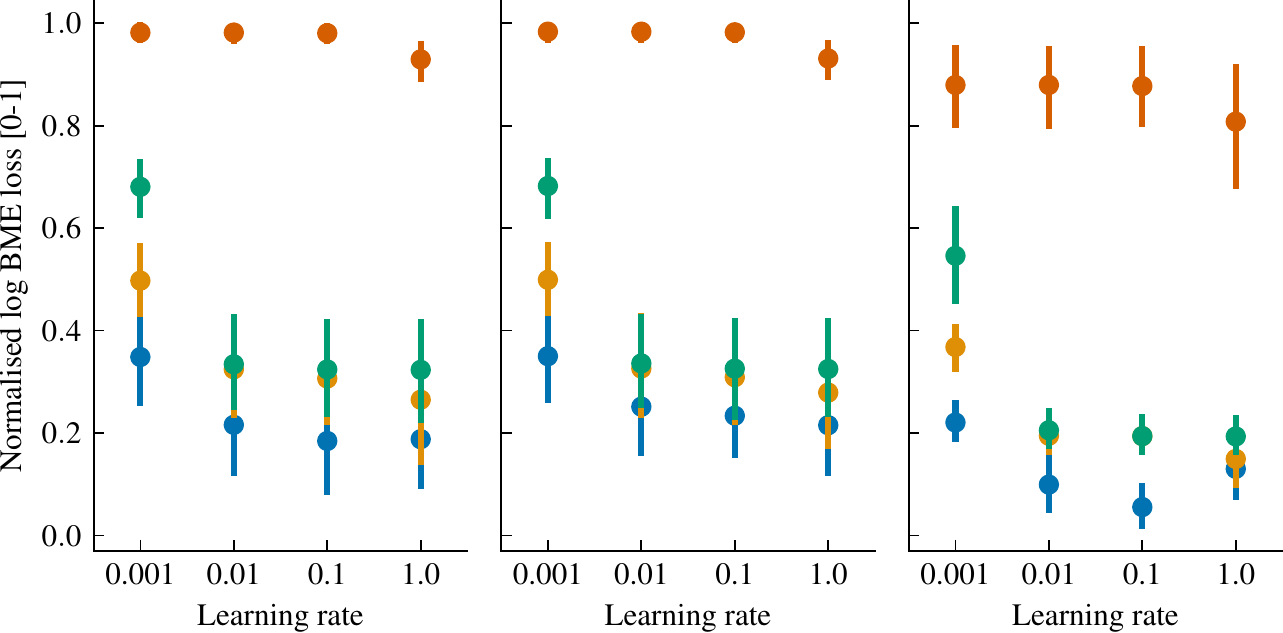}
    \end{subfigure}
    \caption{Convergence analysis of different optimisers and DNA substitution models for the datasets DS1-DS11 (see Table~\ref{tab:data}). \textsb{(a)} Number of steps needed to reach convergence (tolerance: 1e-10). \textsb{(b)} Loss reached at convergence. For each dataset, the log BME losses were scaled using min-max normalisation. Error bars denote 95\% confidence intervals computed with 1000 bootstraps.}
    \label{fig:convergence_analysis}
\end{figure*}

Figure~\ref{fig:convergence_analysis} compares the performance of different optimisers (Adafactor~\parencite{shazeer2018}, AdamW~\parencite{loshchilov2018}, RMSprop~\parencite{tieleman2012}, and SGD) under different DNA substitutions models (JC69~\parencite{jukes1969}, F81~\parencite{felsenstein1981}, TN93~\parencite{tamura1993}) on a single optimisation step (no subsequent reordering with Queue Shuffle) for a maximum of 5000 steps. Four learning rates were considered, logarithmically spaced from 0.001 to 1.0. Whereas convergence speed appears to be independent of the chosen DNA substitution model, the results varied widely with respect to the optimisation algorithm. In particular, Adafactor optimisation produced the best performance, with increasing convergence speed as learning rates were higher.

\subsection{Miscallaneous Lemmas}
\subsubsection*{Supplementary lemmas for Lemma~\ref{lem:root_min}}
\begin{lemma}
\label{lem:supp_1}
    Using the notation of Lemma \ref{lem:root_min}, define $D_{Ar}$ to be the distance between node $A$ and the root, and $g_A$ to be the path length (i.e. the generation of node $A$). Then, 
\begin{equation}
\label{eq:root_reform}
    \sum_{X_b(A,B) = 1}D_{AB}2^{-e_{AB}} = 2\sum_A2^{-g_A}D_{Ar}
\end{equation}
where the $e_{AB}$ terms refer to the path lengths in the original unrooted tree.
\end{lemma}

\textbf{Proof:} We proceed by induction on the number of leaf nodes, $n$. For $n=2$, the root must be placed on the edge connecting nodes $0$ and $1$ and $e_{01} = 1$. Then
\begin{equation}
     \sum_{X_b(A,B) = 1}D_{AB}2^{-e_{AB}} = \frac{1}{2}(D_{01} + D_{10})  =\bigg(2^{1-1}D_{0r} + 2^{1-1}D_{1r}\bigg)
\end{equation}
as required, noting that $D_{0r} + D_{1r} = D_{01}$. Suppose that our claim holds for $n = k$ and consider a tree with $n = k+1$. Choose a pair of sibling leaf nodes (i.e. leaf nodes joined to the same internal node) $y$ and $z$ such that the root is not on an edge joining one of these nodes to the tree (this must exist as, for $n > 2$, there are at least two pairs of sibling leaf nodes). Suppose that $y$ and $z$ are connected to the internal node $w$. Now, note that the contribution to the objective of node $y$ is
\begin{align}
    2\sum_{B:X_b(y,B) = 1}D_{yB}2^{-e_{yB}} &= 2\sum_{B:X_b(w,B) = 1}(D_{wB} + D_{wy})2^{-e_{yB}}\\
    &= 2\sum_{B:X_b(w,B) = 1}D_{wB}2^{-e_{yB}} + 2D_{wy} \sum_{B:X_b(w,B) = 1}2^{-e_{yB}}\\
    \label{eq:final_sum} &= \sum_{B:X_b(w,B) = 1}D_{wB}2^{-e_{wB}} + 2D_{wy} \sum_{B:X_b(w,B) = 1}2^{-e_{yB}}
\end{align}
where the factor of $2$ comes from the fact that both $(y,B)$ and $(B,y)$ must be considered. The final term in this equation can be simplified. Consider the subtree whose leaves are the root and all nodes such that $X_b(w,B) = 1$. Then, by the Kraft Equality on this subtree
\begin{equation}
    \sum_{B:X_b(w,B) = 1}2^{-g_B} =\frac{1}{2}
\end{equation}
as $g_B$ gives the distance between the root and node $B$. Moreover, as $e_{yB} = g_y + g_B - 1$ (where the $-1$ accounts for the fact that $\mathcal{T}$ is unrooted)
\begin{equation}
    \sum_{B:X_b(w,B) = 1}2^{-e_{yB}} =  2^{1-g_y}\sum_{B:X_b(w,B) = 1}2^{-g_B} = 2^{-g_y}
\end{equation}
Thus, (\ref{eq:final_sum}) becomes
\begin{equation}
   2 \sum_{X_b(y,B) = 1}D_{yB}2^{-e_{yB}} =  \sum_{B:X_b(w,B) = 1}D_{wB}2^{-e_{wB}} + 2D_{wy}2^{-g_y}
\end{equation}
Thus, the change to $\sum_{X_b(A,B) = 1}D_{AB}2^{-e_{AB}} $ caused by adding nodes $y$ and $z$ from the tree is to subtract
\begin{equation}
    2\sum_{X_b(w,B) = 1}D_{wB}2^{-e_{wB}}
\end{equation}
(as node $w$ is no longer a leaf node) and to add the contributions from nodes $y$ and $z$
\begin{equation}
    2\bigg(\sum_{B:X_b(y,B) = 1}D_{yB}2^{-e_{yB}}  +  \sum_{B:X_b(z,B) = 1}D_{zB}2^{-e_{zB}}\bigg)  =  \sum_{X_b(w,B) = 1}D_{wB}2^{-e_{wB}} + 2D_{wy}2^{-g_y} + \sum_{X_b(w,B) = 1}D_{wB}2^{-e_{wB}} + 2D_{wz}2^{-g_z}
\end{equation}
There is no $(y,z)$ term to consider as $X_b(y,z) = 0$. Note that the $\sum_{X_b(w,B) = 1}D_{wB}2^{-e_{wB}}$ terms cancel. Define   $\mathcal{T}'$ to be the tree when node $y$ and $z$ are removed so, as it now has $k$ leaf nodes, the inductive hypothesis can be used to show
\begin{align}
    \sum_{A,B\in \mathcal{T}:X_b(A,B) = 1}D_{AB}2^{-e_{AB}} &=  \sum_{A',B' \in \mathcal{T}':X'_b(A',B') = 1}D_{A'B'}2^{-e'_{A'B'}} + 2D_{wy}2^{-g_y} + 2D_{wz}2^{-g_z}\\
    &= 2\sum_{A' \in \mathcal{T}'}2^{-g'_{A'}}D_{A'r'}+ 2D_{wy}2^{-g_y} + 2D_{wz}2^{-g_z}\\
     &= 2\sum_{A' \in \mathcal{T}'/\{w\}}2^{-g'_{A'}}D_{A'r'}+ 2D_{wr}2^{-g_w}+ 2D_{wy}2^{-g_y} + 2D_{wz}2^{-e_{zr}} 
\end{align}
where dashes are used to denote quantities in $\mathcal{T}'$. Note that 
\begin{equation}
    \sum_{A' \in \mathcal{T}'/\{w\}}2^{-g'_{A'}}D_{A'r'} = \sum_{A \in \mathcal{T}/\{y,z\}}2^{-g_{A}}D_{Ar}
\end{equation}
as only the nodes $w$, $y$ and $z$ are affected by changing from $\mathcal{T}$ to $\mathcal{T}'$. Moreover, note that
\begin{equation}
    2^{-g_w} = 2^{-g_y} + 2^{-g_z}
\end{equation}
as $g_w = g_y-1 = g_z-1$, and, as the distances are additive,
\begin{equation}
    D_{wy} + D_{wr} = D_{yr} \quad \text{and} \quad D_{wz} + D_{wr} = D_{zr}
\end{equation}
Hence
\begin{align}
    2D_{wy}2^{-g_y} + 2D_{wz}2^{-g_z} + 2D_{wr}2^{-g_w} &= 2D_{wy}2^{-g_y} + 2D_{wz}2^{-g_{z}} + 2D_{wr}(2^{-g_y} + 2^{-g_z})\\
    &= 2(D_{wy} + D_{wr})2^{-g_y} + 2(D_{wz} + D_{wr})2^{-g_z}\\
    &=2D_{yr}2^{-g_y} + 2D_{zr}2^{-g_z}
\end{align}
Thus,
\begin{equation}
    \sum_{X_b(A,B) = 1}D_{AB}2^{-e_{AB}} =\sum_A2^{-g_A}D_{Ar}
\end{equation}
and hence the claim holds by induction as required.
\begin{lemma}
\label{lem:supp_2}
    Using the notation of Lemma \ref{lem:root_min},
    \begin{equation}
    \mathcal{D} = \sum_A2^{-g_A}D_{Ar}
\end{equation}
\end{lemma}
\textbf{Proof:} This can be proved through induction on the number of nodes. It clearly holds if there are only two nodes in the tree. Assume it holds whenever there are $k$ nodes, and consider a tree with $k+1$ nodes. We process the first pair of leaf nodes, $y$ and $z$, and assume that their associated distances to their parent, $w$ are $D_{wy}$ and $D_{wz}$ and that the new tree created is $\mathcal{T}'$ with distances $D'$. Then, we have, by our inductive hypothesis
\begin{align}
   \mathcal{D} &=   \sum_{A' \in \mathcal{T}'}2^{-g'_{A'}}D'_{A'r'}\\
   &= \sum_{A' \in \mathcal{T}'/\{w\}}2^{-g'_{A'}}D'_{A'r'} + 2^{-g'_w}D'_{wr'}\\
   &= \sum_{A \in \mathcal{T}/\{y,z\}}2^{-g_A}D_{Ar} + 2^{-g_w}\bigg(D_{wr} + \frac{D_{wy} + D_{wz}}{2}\bigg)\\
    &= \sum_{A \in \mathcal{T}/\{y,z\}}2^{-g_A}D_{Ar} + 2^{-g_w}\bigg(\frac{D_{yr} + D_{zr}}{2}\bigg)\\
     &= \sum_{A \in \mathcal{T}}2^{-g_A}D_{Ar}
\end{align}
as required, where we have used the fact that $D_{wr} + D_{wy} = D_{yr}$ and that $2^{-g_w} = 2\times2^{-g_y} =2\times2^{-g_z}$.
\subsubsection*{Supplementary Lemmas for Lemma \ref{lem:swap}}
\begin{lemma}
\label{lem:supswap1}
   Using the notation of Lemma \ref{lem:swap}, for a given permutation $\sigma$,
\begin{equation}
\label{eq:equallike}
    \mathbb{P}\bigg[Q(\mathcal{T}) = \sigma\bigg] \in \bigg\{0,\frac{1}{2^{n-1}}\bigg\}
\end{equation}
\end{lemma}
\textbf{Proof} Suppose that every (internal and external) node in $\mathcal{T}$ is assigned a fixed unique position, such that one can consistently define a “leftmost'' and “rightmost'' node from a pair of nodes. Then, the randomness in the Queue Shuffle algorithm can be represented by a uniform random vector $\boldsymbol{r} \in \{0,1\}^{n-1}$ such that when the $m^{\text{th}}$ internal node is processed, the leftmost child is given the same label as its parent if and only if $r_{m+1} = 1$. Each distinct $\boldsymbol{r}$ results in a distinct ordering (as, given the leaf labels, one can generate the unique labelling of the tree as a parent must have a label equal to the minimum of the labels of its two children). Hence, each possible labelling is equally likely, giving (\ref{eq:equallike}) and completing the proof.
\begin{lemma}
\label{lem:supswap2}
Using the notation of Lemma \ref{lem:swap}
    \begin{equation}
    \mathbb{P}\bigg[l^{\mathcal{S}}_j({Q(\mathcal{T})})  =l_i({Q(\mathcal{T})})  \bigg] = \frac{1}{4}
\end{equation}
Moreover, $l^{\mathcal{S}}_j({Q(\mathcal{T})})  =l_i({Q(\mathcal{T})})$ if and only if node $j$ was placed ahead of the root of $\mathcal{S}_i$ in the queue
\end{lemma}
\textbf{Proof} Using the $\boldsymbol{r}$ defined in the proof of Lemma \ref{lem:supswap1}, define the random indices $A$ and $B$ such that $r_A$ corresponds to processing node $i$ and $r_B$ corresponds to processing node $j$. Note that $r_A$ and $r_B$ are independent (though $B$ will in general depend on $r_A$). 
\\
\\
\noindent
Now, suppose without loss of generality that $j$ is the leftmost child of $i$ and that the root of $\mathcal{S}_j$ is the leftmost child of $j$. As children have labels greater than or equal to their parents, 
\begin{equation}
     \mathbb{P}\bigg[l^{\mathcal{S}}_j({Q(\mathcal{T})})  =l_i({Q(\mathcal{T})})\bigg] = \mathbb{P}\bigg[l^{\mathcal{S}}_j({Q(\mathcal{T})})  = l_j({Q(\mathcal{T})})  =l_i({Q(\mathcal{T})})\bigg]
\end{equation}
and hence
\begin{equation}
     \mathbb{P}\bigg[l^{\mathcal{S}}_j({Q(\mathcal{T})})  =l_i({Q(\mathcal{T})})\bigg] = \mathbb{P}(r_A = r_B = 1) = \frac{1}{4}
\end{equation}
which is the first result. The second result follows immediately from the fact that $r_A = 1$.

\subsection{Estimation of GTR+\texorpdfstring{$\Gamma$}~~distances}
\label{append:est}
To estimate distances under a GTR+$\Gamma$ substitution model we use the approach outlined in~\parencite{yang2006}. Assuming a general time reversible rate symmetric matrix $Q$, the transition probability matrix over time is found via the matrix exponential $P(t)=e^{Qt}$. This matrix exponential can be readily computed via eigendecomposition. 
\\
\\
\noindent
Including variable rates among sites for a Gamma distribution $g$, $P(t)=\int e^{Qu} g(u) du$, which can again be estimated via eigendecomposition. Given parameters for rates, $S$, frequencies, $\pi$, the time between two sequences $t_{ij}$, and genetic sequence data $\mathcal{G}$, the log-likelihood for the transitions between a pair of taxa $i$ and $j$ is
\begin{equation}
    \mathcal{L}_{ij}(\mathcal{G}|S,\pi,t_{ij}) = \sum_a \sum_b \kappa^{ij}_{ab} \log(P_{ab}(t_{ij}; S,\pi))
\end{equation}
where $\kappa^{ij}_{ab}$ is the number of $a \to b$ transitions from taxon $i$ to taxon $j$. We approximate the optimal parameters by maximizing the total log-likelihood (that is, the sum over $i$ and $j$ of $\mathcal{L}_{ij}$) using gradient descent in Jax.
\subsection{Fast discrete hill-climbing with Phylo2Vec}
The computational complexity of GradME is substantially higher than that of FastME due to the continuous nature of the algorithm. Thus, particularly for large numbers of taxa, using a similarly fast algorithm, at least to get close to the optimal tree, may be preferable.

Because of this, we have developed an alternative, discrete algorithm which has the same computational complexity as FastME. At each step, our algorithm outputs a matrix $\Delta$ such that $\Delta_{ij}$ is the change in the objective function if the value of $v_i$ were changed to be equal to $j$. $\Delta$ allows us to perform a hill-climbing optimisation, as the change corresponding to the minimum value of $\Delta$ is made.

Analogously to FastME, $\Delta$ can be calculated in $\mathcal{O}(n^2\text{diam}(\mathcal{T}))$ time, where $\text{diam}(\mathcal{T})$ is the maximum inter-taxa path length of the tree (this is generally substantially smaller than $n$). This algorithm works exclusively for unrooted trees, though a similar algorithm could be developed using our rooted objective.

Motivated by the method of \parencite{Desper2002-uy}, our algorithm begins by calculating directed edge weight vectors $\boldsymbol{w}_e^{\pm}$ for the edges in $\mathcal{T}$. Each edge $e$ naturally partitions $\mathcal{T}$ into two disjoint subtrees, $\mathcal{S}_1^e$ and $\mathcal{S}_2^e$, with each one being rooted at a node connected to $e$. We suppose that $\mathcal{S}_1^e$ is the tree not containing some fixed node $X$. Then, we assign $\boldsymbol{w}_e^{+}$ to be the balanced distance between the root of $\mathcal{S}_1^e$ and each of the individual leaf nodes in $\mathcal{S}_2^e$, and  $\boldsymbol{w}_e^{-}$ to be the balanced distance between the root of $\mathcal{S}_2^e$ and each of the individual leaf nodes in $\mathcal{S}_1^e$. These edge weights can be calculated efficiently by using an iterative scheme, using the fact that the weight on a given edge from an internal node $x$ to another node $y$ is the mean of the weights of the two edges from nodes $z$ and $a$ (both distinct from $y$) to $x$ (recalling that these weights are directional). This iterative scheme is closed by the fact that the weights on edges from leaf nodes to the tree are given by the appropriate columns of the distance matrix $D$.

This weighted tree can be used to calculate the distance between any pair of subtrees. For each fixed subtree, $\mathcal{S}$, one can iteratively calculate the distance $d_{\mathcal{S} \mathcal{R}}$ between it and (disjoint) subtrees $\mathcal{R}$, using the fact that the precalculated $\boldsymbol{w}$ terms give the distance between each subtree and each leaf, and that for any pair of subtrees $\mathcal{R}_1$ and $\mathcal{R}_2$ which share a parent node and can therefore be combined in a subtree $\mathcal{R}_3$,
\begin{equation}
    d_{\mathcal{S} \mathcal{R}_3} = \frac{1}{2}\bigg(d_{\mathcal{S} \mathcal{R}_1}+d_{\mathcal{S} \mathcal{R}_2}\bigg)
\end{equation}
From \parencite{Desper2002-uy}, there exists a simple formula for the difference in objective function caused by pruning a subtree and regrafting it to an adjacent edge. If an internal node $x$ is joined to nodes which are the roots of disjoint subtrees $A$, $B$ and $C$, then the difference between attaching a subtree $K$ to the edge joining $x$ and $C$ and attaching the subtree $K$ to the edge joining $x$ and $B$ is
\begin{equation}
   \frac{1}{4}\bigg( \delta_{AB} + \delta_{KC} - \delta_{AC} - \delta_{KB}\bigg)
\end{equation}
where here, $\delta$ denotes inter-subtree distance. These $\delta$ terms are not equivalent to the $d$ terms, as removing $K$ from the original tree creates a different set of possible subtree pairs. However, they can be calculated in $\mathcal{O}(n^2\text{diam}(\mathcal{T}))$ time from the distances $d$ following the methods of \parencite{Desper2002-uy}.

Explicitly, our algorithm moves outwards from the parent of the subtree which we are removing. We take $B$ to be the node we are currently processing, $x$ to be the previously processed node on this path, $C$ to the node processed two iterations ago, and $A$ to be the other node connected to $x$. Then, one has simply
\begin{equation}
    \delta_{AB} = d_{AB} \quad \text{and} \quad \delta_{KB} = d_{KB}
\end{equation}
Calculating $\delta_{AC}$ is slightly more complicated as the subtree $K$ has been removed from $C$. If the original root of $K$ was a path length $l$ from $C$, and if that root shares a parent node (in $C$) with a subtree $F$, then
\begin{equation}
    \delta_{AC} = d_{AC} + 2^{-l}(d_{AF} - d_{AK})
\end{equation}
as the subtree $F$ now has double the weight in $C$. Finally, to calculate $d_{CK}$, one must calculate the distance between $K$ and all subtrees connected to, but disjoint from the path taken from the tree root to $C$. If these subtrees are $\mathcal{S}_1,...,\mathcal{S}_m$ respectively and are distance $1,...,m$ from $C$, then
\begin{equation}
    \delta_{KC} = \sum_{q = 1}^m 2^{-q}d_{C \mathcal{S}_q}
\end{equation}
It is this step which pushes the complexity of our algorithm above $n^2$

Thus, one can calculate the objective differences caused by each subtree-prune and regraft move by repeatedly applying this formula for each pruned subtree, essentially ``moving'' this subtree around the remaining tree. 

The final step of our algorithm is to find the corresponding subtree-prune and regraft move for each possible change of $\boldsymbol{v}$. When $v_i$ is changed to be equal to $j$, the subtree that is moved is the largest subtree of $\mathcal{T}$ containing $i$ such that the leaves all have labels greater than or equal to $i$ (in many cases, this will simply be the node $i$). For each value of $i$, one can find the edges which were connected to each leaf node when node $i$ was attached in the left-to-right construction algorithm. The edge connected to node $j$ at this stage gives the location of the regraft position of the subtree.

\clearpage
\subsection{Code: continuous BME objective function}
\label{append:code}
\begin{minted}[fontsize=\fontsize{5.5}{7.5}\selectfont]{python} 
import jax.numpy as np

from jax import jit, lax
from jax.scipy.special import logsumexp

@jit
def get_edges_exp_log(W, rooted):
    """Calculate the log-expectation of the objective value of a tree drawn with distribution W.
    We calculate and update E_ij throughout the left-to-right construction procedure

    Args:
        W (jax.numpy.array): Tree distribution
        rooted (bool): True is the tree is rooted, otherwise False

    Returns:
        E (jax.numpy.array): Log of the expected objective value of a tree drawn with W
    """
    # Add jnp.finfo(float).eps to W.tmp to avoid floating point errors with float32
    W_tmp = (
        jnp.pad(W, (0, 1), constant_values=jnp.finfo(float).eps) + jnp.finfo(float).eps
    )

    n_leaves = len(W) + 1

    E = jnp.zeros((n_leaves, n_leaves))
    E = E.at[1, 0].set(
        0.5 * E[0, 0] * W_tmp[0, 0] + jnp.log(0.25 * (2 - rooted) * W_tmp[0, 0])
    )
    E = E + E.T

    trindx_x, trindx_y = jnp.tril_indices(n_leaves - 1, -1)

    def body(carry, _):
        E, i = carry

        E_new = jnp.zeros((n_leaves, n_leaves))

        trindx_x_i = jnp.where(trindx_x < i, trindx_x, 1)
        trindx_y_i = jnp.where(trindx_x < i, trindx_y, 0)

        indx = (trindx_x_i, trindx_y_i)

        E_new = E_new.at[indx].set(
            E[indx] + jnp.log(
                1 + jnp.finfo(float).eps - 0.5 * (W_tmp[i - 1, indx[1]] + W_tmp[i - 1, indx[0]]) 
            )
        )

        # exp array
        mask_Ei = jnp.where(jnp.arange(n_leaves) >= i, 0, 1)
        exp_array = E * jnp.where(jnp.arange(n_leaves) >= i, 0, mask_Ei.T)

        # coef array
        mask_Wi = jnp.where(jnp.arange(n_leaves) >= i, 0, 0.5 * W_tmp[i - 1])
        coef_array = (jnp.zeros_like(W_tmp) + mask_Wi).at[:, i].set(0.25 * W_tmp[i - 1])
        coef_array = coef_array * (1 - jnp.eye(W_tmp.shape[0]))

        # logsumexp
        tmp = logsumexp(exp_array, b=coef_array, axis=-1) * mask_Ei

        E_new = E_new.at[i, :].set(tmp)

        # Update E
        E = E_new + E_new.T

        return (E, i + 1), None

    (E, _), _ = lax.scan(body, (E, 2), None, length=n_leaves - 2)

    return E

@jit
def bme_loss_log(W, D, rooted):
    """Log version of the BME loss function

    Args:
        W (jax.numpy.array): Tree distribution
        D (jax.numpy.array): Distance matrix
        rooted (bool): True is the tree is rooted, otherwise False

    Returns: 
        loss (float): BME loss
    """
    E = get_edges_exp_log(W, rooted)
    loss = logsumexp(E, b=D)
    return loss
\end{minted}

\end{document}